\documentclass{lmcs}

\usepackage{amsthm}
\usepackage{amssymb}
\usepackage{amsfonts}
\usepackage{mathrsfs}
\usepackage{multicol} 
\usepackage{array} 
\usepackage{commath}

\input xy
\xyoption{all}

\newcommand{\st}{\text{ }|\text{ }}

\newcommand{\R}{\mathbb{R}}

\newcommand{\meas}{\mathbf{Unc} }
\newcommand{\ing}{\mathsf{Ing }\ }

\newcommand{\pr}{\mathsf{pr}}
\newcommand{\inc}{\mathsf{in}}
\newcommand{\ev}{\mathsf{ev}}
\newcommand{\sig}{\mathsf{next}}
\newcommand{\ded}{\vdash}
\newcommand{\sat}{\models}
\newcommand{\class}[1]{[\![#1]\!]} 
\newcommand{\ere}{\mathsf{r}}
\newcommand{\bbel}{\mathsf{Bel}} 

\newcommand{\deltapl}{\Delta_{Pl}} 
\newcommand{\plaus}{\gamma}
\newcommand{\bel}{\underline{\gamma}} 
\newcommand{\pplaus}{\mathsf{Plaus}}
\newcommand{\poss}{\eta} 
\newcommand{\nec}{\underline{\eta}} 
\newcommand{\pposs}[1]{{\sf Ps}_{#1}} 
\newcommand{\nnec}[1]{{\sf Nc}_{#1}}
\newcommand{\upper}[1]{{\sf U}_{#1}}
\newcommand{\plower}[1]{{\sf L}_{#1}}

\newcommand{\pl}[1]{{\sf Pl}_{#1}}
\newcommand{\proba}[1]{{\sf P}_{#1}}
\newcommand{\bl}[1]{{\sf Bf}_{#1}}
\newcommand{\interp}[1]{\|{#1}\|}
\newcommand{\vabs}[1]{\left|{#1}\right|}

\makeatletter
\renewcommand{\p@enumii}{}
\makeatother

\begin{document}
	\title{Coalgebraic Modal Logic for Dynamic Systems with Uncertainty}

	\author[A.~Gallardo]{Andr\'es Gallardo\lmcsorcid{0009-0009-3516-3014}}
	\author[I.Viglizzo]{Ignacio Viglizzo\lmcsorcid{0000-0002-5303-623X}}
	
	\address{Instituto de Matemática (INMABB), Departamento de Matemática, Universidad Nacional del Sur (UNS) -CONICET, Bahía Blanca, Argentina}	
	\email{andresgallardo123@gmail.com , viglizzo@criba.edu.ar}  


	\numberwithin{equation}{section}

	\begin{abstract}
		In this paper we define a class of polynomial functors suited for constructing coalgebras representing processes in which uncertainty plays an important role. In these polynomial functors we include upper and lower probability measures, finitely additive probability measures, plausibilty measures (and their duals, belief functions), and possibility measures. We give axioms and inference rules for the associated system of coalgebraic modal logic, and construct the canonical coalgebras to prove a completeness result.
	\end{abstract}
	\maketitle

	\section{Introduction}
	
	Coalgebras are used as a common abstraction of different state-based dynamic systems of which some behavior can be observed, like labeled transition systems, automata, Markov chains, and Kripke frames. Given a category $\bf C$ and an endofunctor $T$ on $\bf C$, a $T$-{\em coalgebra} is defined as a pair $(X,c)$ where $X$ is an object of the category and $c$ is a morphism from $X$ to $TX$. The fact that Kripke frames can be written as coalgebras inspired the definition of coalgebraic modal logic for a functor in general \cite{moss99coalgebraic}. Real-world systems evolve, transitioning from one state to another, and coalgebras provide a framework for capturing these dynamics and a natural bridge to construct logics for those systems.
	
	In a category with binary products and coproducts, one can define a class of {\em polynomial functors}: taking the identity functor and the constant functor for each object in the category as polynomial functors, if $T_1$ and $T_2$ are polynomial, so are the functors $T_1\times T_2$ and $T_1+T_2$. Thus a functor like $T(X)=X\times X+ X$ is a polynomial functor\footnote{There is a different notion of polynomial functor in linear algebra. The polynomial functors we mention here can be seen as a particular case of the more general ones defined in \textit{type theory}.}.  For the particular cases of the polynomial functors in the category of sets, R\"o\ss iger gave in \cite{rossiger00coalgebras} a modular way of defining a coalgebraic modal logic tailored to describe, and reason about, the observations in coalgebras for these polynomial set functors. This modal logic captures the inherent dynamics of the system.
	
	Moss and Viglizzo in \cite{moss04harsanyi} extended the definition of polynomial functors to the category of measurable spaces, by adding the closure under a functor $\Delta$ that when applied to a measurable space $X$ gives the space $\Delta X$ of all the probability distributions over $X$. In that work, they constructed the final coalgebras for these functors using sets of satisfied formulas. As an application, they presented the construction of Harsanyi type spaces as final coalgebras for a polynomial functor in the category of measurable spaces.  Harsanyi types are descriptions of players of a game based on their beliefs, where the beliefs are modeled as probability measures on the sets of the types of the rest of the players and other unknowns. Other applications of polynomial measure functors include Markov chains, probabilistic automata, probabilistic labeled transition systems, and Fagin and Halpern's probability frames \cite{fagin94reasoning} (see Example \ref{exampleProbabilisticFrames} below). Further examples of coalgebras with probabilistic components can be found in \cite{sokolova11probabilistic}. 
	
	Later on, Friggens and Goldblatt gave in \cite{goldblattfriggens06} a deduction theory for the formulas of coalgebraic modal logic that gives a syntactic way of constructing the final coalgebras and proved a completeness result. A deduction system with axioms and rules of inference adds valuable structure, making some assumptions in the modeling more explicit, allowing for automation  and providing a versatile framework that can be applied to different areas by changing the underlying axioms.  Goldblatt expanded in \cite{goldblattdeduction2010} his work with Friggens to encompass the measurable spaces of probability distributions as well.
	
	Using a different approach, Jacobs and Sokolova \cite{jacobs10exemplaric} explored the expressivity of a coalgebraic modal logic for some specific functors (including $\Delta$) through properties of the ambient category and the dual equivalence between coalgebras for those functors and meet semi-lattices. Their work provides valuable insights within this framework. However, for this paper, we go in a different direction, developing a deduction theory for a wider class of polynomial functors dealing with different representations of uncertainty.
	
	Uncertainty can easily arise in mathematical models, and it has many sources, like errors in measurement, changing conditions, or lack of information. In the face of it, an agent must resort to forming their beliefs, through hopefully educated guesses. We will focus on four different representations of uncertainty, working in the category $\meas$ of sets endowed with a boolean algebra of measurable subsets, and measurable functions between them.   We call these spaces \textit{uncertainty spaces}, following \cite{galeazzi23choice}. 
	
	The most common way of representing uncertainty is to use probability distributions, in what is known as the bayesian approach, starting with the work by de Finetti \cite{deFinetti31significato}. Admittedly, it is hard to pin down an exact number to the likelihood of each considered event (i.e. measurable subset). One way of dealing with this problem is to consider more than one plausible probability distribution and then use, as a basis for reasoning, the supremum and infimum of the probabilities that these measures assign to the event. This defines the notion of upper and lower probabilities, thus providing a range of possible probabilities, and were first considered by Dempster in \cite{dempster67upper}, although the idea of using bounds for the unknown probability of an event, dates back to Boole (\cite{Boole1854thought}), in his theory of logical equations for probabilities. Upper and lower probabilities can also be regarded as a special kind of interval-valued probabilities (see \cite{Weichselberger00interval}). 
	
	Other approaches we consider are  Dempster-Shafer belief functions and possibility measures.  Dempster-Shafer belief functions \cite{shafer76evidence} instead of directly assigning probabilities to events, assign \textit{belief mass} to subsets of the power set of possible outcomes and then calculate the likelihood of an event by adding the masses contained in that event. There is a {\em rule of combination} that provides a way of integrating differing belief functions, and this propitiates the application of these functions to statistical inference and decision-making, for example combining information from multiple sensors with varying levels of reliability and incorporating information from various diagnostic tests \cite{smets99practical}.

	Possibility measures, in turn, provide a way to express the degree of possibility or plausibility of events in a fuzzy set framework \cite{zadeh78possibility}, based on the premise that the maximum value of the possibility gives the likelihood of an event measure over the elements in the event.
	
	A wider survey on ways of modeling uncertainty can be found in \cite{halpern17reasoning}. Each approach is suited to different types of uncertainty and is applied in various domains depending on the nature of the problem at hand.
	
	There have been some axiomatic systems of modal logic about probabilities, starting with \cite{LarsenSkou91}, and including \cite{fagin94reasoning}, \cite{goldblattdeduction2010}, \cite{Meier2012inflogic}, and \cite{Baratella2022inflogic}. For upper and lower probabilities such a system was developed by Savi\'c et.al, \cite{dodersavic2015} and \cite{doder17logics}.
	
	In order to obtain axiomatic systems for the polynomial functions, a good characterization of the representation of uncertainty is needed. 
	A characterization of functions that are the supremum of finitely additive functions was given already by Lorentz in \cite{lorentz52multiply}. Other characterizations were given in \cite{williams76indeterminate},  \cite{wolf77obere}, and \cite{angerlembcke85}.  The latter of these characterizations, due to Anger and Lembcke was used by Savić, Doder and Ognjanović in \cite{dodersavic2015} to define a sound and strongly complete modal logic for upper and lower probabilities.
	
	The rest of this paper is structured as follows: in the second section we review the notions of uncertainty spaces, upper probability measures and their characterization, the functor $\Delta^*$, upper probability polynomial functors, their ingredients, and associated multigraphs. 
	
	In the third section, we develop the coalgebraic modal logic for $T$-coalgebras where $T$ is an upper measurable polynomial functor, giving axioms for the corresponding $T$-deduction systems. In the fourth section, we build the canonical coalgebras used to prove the completeness result.
	
	Finally, in the last section, we expand the class of polynomial functors by adding closure under other functors used for representing uncertainty: first, we use the work done on upper probability measures to handle finitely additive probability measures. Then we move on to plausibility measures and their dual, belief functions, captured by the functor $\deltapl$. The last paradigm for uncertainty representation that we treat is the one of possibility measures, introducing the functor $\Delta_{Ps}$. For all these, we prove the functoriality, give axioms and rules that allow the construction of the canonical coalgebra from sets of formulas, proving the completeness result.

	\section{Preliminaries}
	
	In this section we review upper and lower probability measures and build an endofunctor $\Delta^*$ in the category of uncertainty spaces such that the underlying set of $\Delta^*X$ is the collection of all the definable upper probability measures on $X$. This functor will be key in defining the upper measurable polynomial functors. The coalgebras on these functors will model dynamic systems in which uncertainty is represented by upper and lower probability measures.
	
	\begin{defi}\label{algofsets}
		An \emph{algebra of subsets} of a set $X$ is a nonempty class $\Sigma$ of subsets of $X$ such that if $U\in\Sigma$ and $V\in\Sigma$ then $U\cup V\in\Sigma$, and $X\setminus U\in\Sigma$.
	\end{defi}
	
	As a consequence of the definition, the algebra $\Sigma$ contains $X$, the empty set, and is closed under finite unions and intersections. We will denote the relative complement $X\setminus U$ also by $U^c$ when it is not necessary to specify the ambient set $X$. For a family $\mathcal{F}$ of subsets of $X$, the \emph{algebra} generated by $\mathcal{F}$ will be denoted by $\Sigma(\mathcal{F})$, and it is the least algebra containing $\mathcal{F}$. 
	
	\begin{defi}\label{measurablespace}
		An \emph{uncertainty space}, is a pair $(X,\Sigma)$ where $X$ is a set and $\Sigma$ is an algebra of subsets of $X$. We will call the elements of $\Sigma$ \emph{measurable sets}. Given uncertainty spaces $(X,\Sigma)$ and $(Y,\Sigma')$, a function $f:X\to Y$ is said to be \emph{measurable} if for every $U\in\Sigma'$, $f^{-1}(U)\in\Sigma$. We will denote with $\meas$ the category whose objects are uncertainty spaces and its morphisms are measurable functions.
	\end{defi}
	
	\begin{defi}\label{defmedprobfinita}
		A \emph{finitely additive probability measure} (from now on, probability measure) over an algebra of subsets ($X$,$\Sigma$) is a finitely additive function $\mu:\Sigma\to[0,1]$ such that $\mu(\emptyset)=0$ and $\mu(X)=1$. Being finitely additive means that for disjoint sets $U, V\in\Sigma, \mu(U\cup V)=\mu(U)+\mu(V)$.
	\end{defi}
	
	\begin{defi}	\label{defmedprobsup}
		Given a family $\mathcal{P}=\{\mu_i\}_{i\in I}$, of probability measures over an uncertainty space $(X,\Sigma)$, we can define for each $U\in\Sigma$ the \emph{upper} and \emph{lower} probability for $\mathcal{P}$ respectively as follows:
		\[\mathcal{P}^*(U)=\displaystyle\sup_{\mu_i\in\mathcal{P}}\{\mu_i(U)\},\]
		\[\mathcal{P}_*(U)=\displaystyle\inf_{\mu_i\in\mathcal{P}}\{\mu_i(U)\}.\]
		These measures are called \emph{multiply subadditive functions} in \cite{lorentz52multiply}, and \emph{upper envelopes} in \cite{halpernfagin92belief}.
	\end{defi}
	
	Upper and lower probability measures have the following properties:
	
	\begin{lem}\label{upperprobproperties}
		For a set $\mathcal{P}$ of probability measures defined over an uncertainty space $(X,\Sigma)$, the following properties hold:
		\begin{enumerate}[\normalfont(1)]
			\item $\mathcal{P}^*(\emptyset)=0$, $\mathcal{P}^*(X)=1$, $\mathcal{P}_*(\emptyset)=0$, and $\mathcal{P}_*(X)=1$.\label{upperpropemptyanduniverse}
			\item \underline{Subadditivity:} if $U,V\in\Sigma$ are such that $U\cap V=\emptyset$, $\mathcal{P}^*(U\cup V)\le\mathcal{P}^*(U)+\mathcal{P}^*(V)$.\label{upperpropsubaditivity}
			\item \underline{Superaditivity:} if $U,V\in\Sigma$ are such that $U\cap V=\emptyset$, $\mathcal{P}_*(U\cup V)\ge\mathcal{P}_*(U)+\mathcal{P}_*(V)$.\label{upperpropsuperaditivity}
			\item \underline{Monotonicity:} if $U,V\in\Sigma$, and $U\subseteq V$, $\mathcal{P}^*(U)\le\mathcal{P}^*(V)$ and $\mathcal{P}_*(U)\le\mathcal{P}_*(V)$.\label{upperpropmonotonocity}
			\item \underline{Duality:} $\mathcal{P}^*(U)=1-\mathcal{P}_*(U^c)$.\label{upperpropduality}
		\end{enumerate}
	\end{lem}
	
	In order to give sets of axioms for systems with upper and lower probability measures, we need to characterize them, and the properties in the previous lemma are not enough. We need the following notion of \emph{covering} of a set.
	
	\begin{defi}\label{nkcover} Let $J_m$ be the set $\set{1,\ldots,m}$.
		A set $U$ is said to be \emph{covered} $n$ times by a finite sequence $U_1,\ldots,U_m$ of elements of $\Sigma$ if every element of $U$ appears in at least $n$ sets from $U_1,\ldots,U_m$, that is, for all $x\in U$, there exist different indices $i_1,\ldots,i_n$ in $J_m$ such that for all $j\le n$, $x\in U_{i_j}$. An alternative way of saying this is that 
		\[U\subseteq\bigcup_{\substack{I\subseteq J_m,\\|I|=n}}\Big(\bigcap_{i\in I}U_i\Big).\]
		An $(n,k)$-\emph{cover} of $(U,X)$ is a sequence $U_1,\ldots,U_m$ that covers $X$ $k$ times and covers $U$ $n+k$ times.
	\end{defi}
	
	Like Halpern and Pucella \cite{halpernpucella2002}, we use Anger and Lembcke's \cite{angerlembcke85} characterization of upper and lower probability measures as strictly subadditive functions of order infinity, that is, functions that satisfy the condition \ref{30032024-3} below:
	
	\begin{thmC}[\cite{halpernpucella2002}]\label{caracupprob}
		Let $X$ be a set, $\Sigma$ an algebra of subsets of $X$, and $g:\Sigma\to[0,1]$ a function. There exists a set $\mathcal{P}$ of probability measures with $g=\mathcal{P}^*$ if and only if $g$ satisfies the following three properties:
		\begin{quote}
			\begin{enumerate}[\normalfont(UP1)]
				\item $g(\emptyset)=0$,\label{30032024-1}
				\item $g(X)=1$,\label{30032024-2}
				\item For all non-negative integers $m,n,k$ and all $(n,k)$-covers $U_1,\ldots,U_m$ of $(U,X)$,
				\[
				k+ng(U)\le\displaystyle\sum_{i=1}^mg(U_i).
				\]\label{30032024-3}
			\end{enumerate}
		\end{quote}
	\end{thmC}
	
	\begin{obs}\label{notacionconvencion}
		By Theorem $\rm{\ref{caracupprob}}$, if $g$ is a function that satisfies \ref{30032024-1}, \ref{30032024-2}, and \ref{30032024-3}, then $\underline{g}$ defined for all $U\in\Sigma$  by $\underline{g}(U)=1-g(U^c)$  is a lower probability due to Lemma $\rm{\ref{upperprobproperties}\ref{upperpropduality}}$. Thus, lower probabilities can be characterized as functions that satisfy \ref{30032024-1}, \ref{30032024-2}, and the following condition: For all non-negative integers $m,n,k$ and all $(n,k)$-covers $U_1,\ldots,U_m$ of $(U,X)$,
		\[
		k+n\underline{g}(U)\ge\displaystyle\sum_{i=1}^m\underline{g}(U_i).
		\]
	\end{obs}
	
	\begin{obs}\label{subaditividaddenkcover}
		If $U$ and $V$ are disjoint sets in $\Sigma$, then $U,V$ is a $(1,0)$-cover of $(U\cup V,X)$. Therefore, if $g$ satisfies \ref{30032024-3}, it follows that $g(U\cup V)\le g(U)+g(V)$, that is, we can directly deduce the subadditivity of $g$. 
	\end{obs}
	
	\begin{defi}\label{defbetapq}
		For each uncertainty space $(X,\Sigma)$, we consider the space formed by the set $\Delta^*X$ of all upper probabilities defined over $(X,\Sigma)$, and the algebra  generated by the sets of the form $\beta^{p,q}(U)$, where $p,q\in [0,1]$, $U$ is a measurable subset of $X$ and 
		\[
		\beta^{p,q}(U)=\{g\in\Delta^*X\st g(U)\ge p, 1-g(U^c)\ge q\}.
		\]
	\end{defi}
	
	\begin{lem}\label{deltademedidaesmedida}
		If $(X,\Sigma)$ and $(Y,\Sigma')$ are uncertainty spaces, $f:X\to Y$ is a measurable function, and  for  every $g\in\Delta^*X$ and $U\in\Sigma'$, we put 
		\[(\Delta^*f)(g)(U)=g(f^{-1}(U)),\]
		then $(\Delta^*f)(g)$ is an upper probability measure.
	\end{lem}
	
	\begin{proof}
		Due to Theorem \ref{caracupprob}, we just need to prove that $(\Delta^*f)(g)$ satisfies \ref{30032024-1}, \ref{30032024-2} and \ref{30032024-3}.
		\begin{quote}
			\begin{itemize}
				\item [\ref{30032024-1}] $(\Delta^*f)(g)(\emptyset)=g(f^{-1}(\emptyset))=g(\emptyset)=0$.
				\item [\ref{30032024-2}] $(\Delta^*f)(g)(Y)=g(f^{-1}(Y))=g(X)=1$.
				\item [\ref{30032024-3}] Let $U\in\Sigma'$, and $U_1,\ldots,U_m$ a $(n,k)$-cover of $(U,Y)$.
				In order to prove that
				$$k+ng(f^{-1}(U))\le\displaystyle\sum_{i=1}^mg(f^{-1}(U_i)),$$
				it is enough to prove that $f^{-1}(U_1),\ldots,f^{-1}(U_m)$ is an $(n,k)$-cover of $(f^{-1}(U),X)$. Since $f$ is measurable, these inverse images are measurable sets.
				
				Let $x\in f^{-1}(U)$. Then $f(x)=y\in U$. By hypothesis, there exist $\{i_1,\ldots,i_{n+k}\}\subseteq J_m$ such that for all $j\le n+k$, $y\in U_{i_j}$, that is, $x\in f^{-1}(U_{i_j})$. Therefore, $x$ is in (at least) $n+k$ sets in the sequence $f^{-1}(U_1),\ldots,f^{-1}(U_m)$.
				If $x\in X=f^{-1}(Y)$, an analogous argument shows that $x$ is in (at least) $k$ sets in $f^{-1}(U_1),\ldots,f^{-1}(U_m)$.\qedhere
			\end{itemize}
		\end{quote}
	\end{proof}
	
	\begin{obs}
		In Definition \ref{defbetapq}, for the sets $\beta^{p,q}(U)$ we could consider only rational values of $p$ and $q$. This does not change much the results in this work, and we will point out whenever there is a difference. Here we may mention that even tough we have less measurable sets if we restrict the values of $p$ and $q$ to the rationals, the sets $\beta^{p,q}(U)$ are enough to separate any two different upper measures. Since the algebras we are considering are not closed under countable unions or intersections, the sets $\beta^{p,q}(U)$ with irrational values of $p$ and $q$ cannot be obtained this way.
	\end{obs}
	
	\begin{lem}\label{losbetasmedibles}
		If $f:X\to Y$ is a measurable function, then for all measurable sets $U\subseteq Y$ and $p,q\in[0,1]$,
		\[\beta^{p,q}(f^{-1}(U))=(\Delta^* f)^{-1}(\beta^{p,q}(U)).\]
	\end{lem}
	
	\begin{proof}
		\begin{align*}\beta^{p,q}(f^{-1}(U))&=\{g\in\Delta^*X\st g(f^{-1}(U))\ge p, 1-g((f^{-1}(U))^c)\ge q\}\\
			&=\{g\in\Delta^*X\st g(f^{-1}(U))\ge p, 1-g(f^{-1}(U^c))\ge q\}\\
			&=\{g\in\Delta^*X\st (\Delta^*f)(g)(U)\ge p, 1-(\Delta^* f)(g)(U^c))\ge q\}\\
			&=\{g\in\Delta^*X\st (\Delta^* f)(g)\in\beta^{p,q}(U)\}\\
			&=(\Delta^* f)^{-1}(\beta^{p,q}(U)). \qedhere
		\end{align*}
	\end{proof}
	
	Lemma \ref{losbetasmedibles} proves that $\Delta^*f$ is a measurable function since the sets $\beta^{p,q}(U)$ generate the algebra on $\Delta^*Y$. As a consequence, $\Delta^*$ is a well-defined endofunctor in $\meas$.
	
	In the category $\meas$ we can perform the usual constructions of products and coproducts. More precisely, given uncertainty spaces $X_1$ and $X_2$, we take their product $X_1\times X_2$ with the algebra generated by the sets of the form  $U_1\times U_2$  with $U_i$ a measurable subset of $X_i$ for $i=1,2$. Similarly for coproducts $X_1+X_2$, we consider the usual inclusion maps $in_j:X_j\to X_1+X_2$ and the algebra generated by the inclusions $in_j(U_j)$ of the measurable sets $U_j$ of $X_j$ for $j=1,2$.
	It follows by construction that the projections and injections are all measurable functions. Furthermore, these constructions can be extended to functors by performing them pointwise. This is, if $T_1$ and $T_2$ are endofunctors in $\meas$, so are $T_1\times T_2$ and $T_1+T_2$.
	
	We will also use constant functors in the category $\meas$: for each uncertainty space $(M,\Sigma_M)$, the constant functor $M$ is the one that sends each uncertainty space to $M$ and every measurable function to the identity on $M$.
	
	\begin{defi}\label{defmeasurablepolynomialfunctor}
		The class of \emph{upper probability polynomial functors} is the one containing the identity functor on $\meas$, the constant functors and is closed under the operations of taking binary products, coproducts, and applying $\Delta^*$.
	\end{defi}
	
	The fundamental structures in our work are the coalgebras on these upper probability polynomial functors. We first give the general definition of coalgebras.
	
	\begin{defi}\label{defcoalgebra}
		For a given endofunctor $T$ in a category $\bf C$, a $T$-coalgebra is a pair $(X,\alpha)$, where $X$ is an object in the category and $\alpha$ a morphism from $X$ to $TX$. A morphism of $T$-coalgebras from $(X,\alpha)$ to $(Y,\delta)$ is a $\bf C$-arrow $f:X\to Y$ such that $\delta\circ f=Tf\circ\alpha$.
		\[
		\xymatrix{
			X\ar[r]^{\alpha}\ar[d]_{f}	&TX\ar[d]^{Tf}	\\
			Y\ar[r]_{\delta}& TY
		}
		\]
	\end{defi}
	
	We will work with coalgebras on upper probability polynomial functors. The recursive nature of the definition of these functors allows us to get a handle on how to build a modal language that speaks about the structures modeled by the coalgebras. 
	
	\begin{exa} \label{exampleProbabilisticFrames}
		Kripke frames, commonly used in modal logic are given by a set of \emph{worlds} $W$ interconnected by an \emph{accessibility relation}, where each world represents a possible state of affairs. The accessibility relation can be presented as a function from $W$ to $\mathcal{P}(W)$, the set of parts of $W$, so Kripke frames are coalgebras for the functor $\mathcal{P}$.  In probabilistic frames, \cite{fagin94reasoning},  this structure is enriched by associating to each world a probability distribution over the other worlds, and modeled by a coalgebra for the functor $\Delta$.  By replacing probabilities with upper probabilities, we can capture a broader spectrum of uncertainty, encompassing scenarios where precise probabilities are unavailable or insufficient, thus we can study upper probability frames as coalgebras $\alpha:X\to\Delta^*(X)$. 
		
		In the realm of game theory, \emph{type spaces} offer a tool for modeling players' beliefs and uncertainties about their opponents' strategies and the underlying state of the world. Each element of a type space represents a unique type of player, characterized by their beliefs and uncertainties. This interplay is captured by considering coalgebras with maps from a space $X$ (of types) to $\Delta(X\times M)$, \cite{moss04harsanyi}. The coalgebraic structure maps each type to its associated beliefs about other players' types and uncertainties about the world. Again, we can replace $\Delta$ with $\Delta^*$ to allow for a broader representation of uncertainty.
	\end{exa}
	
	Our next goal is to introduce a logic based on each polynomial endofunctor $T$. To do this, we need the following concepts.
	
	Each upper probability polynomial functor $T$ is built up from a set of basic components. We collect these components and the intermediate stages of the construction of $T$ in the set defined below:
	
	\begin{defi}
		If $T$ is an upper probability polynomial functor, the set $\ing T$ of \emph{ingredients} of $T$, is defined by:
		\begin{itemize}
			\item If $T=M$ is a constant functor, or $T=Id$ then $\ing T=\set{T, Id}$,
			\item if $T=T_1+T_2$, or $T=T_1\times T_2$ then $\ing T=\set{T}\cup\ing T_1\cup\ing T_2$, and 
			\item if $T=\Delta^*S$, $\ing T=\{T\}\cup\ing S$.
		\end{itemize}
	\end{defi}
	
	The set of ingredients of $T$ will serve as a set of nodes of a multigraph that will organize the information about the structure of the functor. The labels on the edges of this multigraph will be used to construct the formulas of our modal language.
	
	\begin{defi} The \emph{labelled multigraph} associated to an upper probability polynomial functor $T$ has $\ing T$ as its set of nodes, with labeled edges $S\overset{\kappa}{\rightsquigarrow}S'$ as follows:
		\begin{itemize}
			\item If $S=S_1\times S_2$, we add the edges $S_1\times S_2\overset{\pr_1}{\rightsquigarrow}S_1$ and $S_1\times S_2\overset{\pr_2}{\rightsquigarrow}S_2$.
			\item If $S=S_1+ S_2$, we add the edges $S_1+ S_2\overset{\inc_1}{\rightsquigarrow}S_1$ and $S_1+ S_2\overset{\inc_2}{\rightsquigarrow}S_2$.
			\item For the ingredient $\Delta^*S$, we add the edges $\Delta^*S\overset{(p,q)}{\rightsquigarrow}S$, for $p,q\in[0,1]$. Again, we can choose to limit the values of $p$ and $q$ to the rational numbers.
			\item If $S=Id$, we add the edge $Id\overset{\sig}{\rightsquigarrow}T$.
		\end{itemize}
	\end{defi}
	
	\section{Modal logic for upper  probability polynomial functors}
	\subsection{Formulas and their semantics}
	
	For each upper probability polynomial endofunctor $T$, we will define a language and the formulas in this language. The formulas will be classified in different sorts. The sorts are the ingredients of $T$, as it was done in \cite{rossiger00coalgebras}, \cite{Jacobs01manySorted} for endofunctors in the category $\mathbf{Set}$, and extended to the category $\mathbf{Meas}$ in \cite{moss04harsanyi}. Our contribution here is to define the formulas of sort $\Delta^*S$ when this is one of the ingredients of $T$.
	
	\begin{defi} We write $\varphi:S$ to indicate that a formula $\varphi$ is of sort $S$, where $S\in\ing T$. We will denote with $\mathsf{Form}_S$ the set of all the formulas of sort $S$ for each $S\in\ing T$. The non-modal formulas are formed according to the following rules:
		\begin{itemize}
			\item There is a formula $\bot_S:S$ for each $S\in\ing T$, 
			\item if $\phi:S$ and $\psi:S$, then $\phi\to\psi:S$,
			\item if $S$ is a constant functor for an uncertainty space $(M,\Sigma_M)$, $A\in\Sigma_M$ or $A=\set{m}$, for $m\in M$, then then $A:M$ is a formula.
		\end{itemize}
		
		We indicate that $\varphi::S$ for $S\in \ing T$ if for every constant functor $M\in\ing T$ and  $A:M$ that is a subformula of $\varphi$, we have that $A$ is a measurable set, that is, $A\in\Sigma_M$. We call these formulas \emph{measurable}, and they will be the ones we are most interested in, but we also need the formulas containing singletons to construct a copy of a measurable space out of formulas.
		
		For every edge in the graph of ingredients labelled by $\kappa$, $S\overset{\kappa}{\rightsquigarrow}S'$ we have a modal operator $[\kappa]$ in the following way:
		\begin{itemize}
			\item if $\kappa\ne(p,q)$, and $\varphi:S'$,  $[\kappa]\varphi:S$ is a formula,
			\item if the edge is $\Delta^*S\overset{(p,q)}{\rightsquigarrow}S$, and $\varphi::S$, then $[(p,q)]\varphi:\Delta^*S$ is a formula.
		\end{itemize}
	\end{defi}

	If $\Gamma\subseteq\mathsf{Form}_S$,  $\Gamma::S$ means that $\varphi::S$ for all $\varphi\in\Gamma$. For $\varphi,\psi:S$, we define as usual the abbreviations $\neg\varphi=\varphi\to\bot_S$, $\varphi\lor\psi=\neg\varphi\to\psi$, $\varphi\land\psi=\neg(\neg\varphi\lor\neg\psi)$,  $\varphi\leftrightarrow\psi=(\varphi\to\psi)\land(\psi\to\varphi)$, and $\top_S=\neg\bot_S$.
	
	\begin{defi}\label{defsemantics}
		Given a $T$-coalgebra $(X,\alpha)$, and a formula $\varphi:S$, the \emph{interpretation} of the formula $\varphi$ is the set $\class{\varphi}_S^\alpha\subseteq S(X)$, defined by:
		
		\begin{quote}  
			\begin{multicols}{2}
				\begin{description}
					\item $\class{\bot_S}_S^\alpha=\emptyset$,
					\item $\class{A}_{M}^\alpha=A$,
					\item $\class{\varphi_1\to\varphi_2}_S^\alpha=(S(X)\setminus\class{\varphi_1}_S^\alpha)\cup\class{\varphi_2}_S^\alpha$,
					\item $\class{[\inc_1]\varphi}_{S_1+S_2}^\alpha=in_1(\class{\varphi}_{S_1}^\alpha)\cup in_2(S_2X)$,
					\item $\class{[\inc_2]\varphi}_{S_1+S_2}^\alpha=in_1(S_1X)\cup in_2(\class{\varphi}_{S_2}^\alpha)$,
					\item $\class{[\pr_j]\varphi}_{S_1\times S_2}^\alpha=\pi_j^{-1}\class{\varphi}_{S_j}^\alpha$,
					\item $\class{[\sig]\varphi}_{Id}^\alpha=\alpha^{-1}\class{\varphi}_T^\alpha$,
					\item $\class{[(p,q)]\varphi}_{\Delta^*S}^\alpha=\beta^{p,q}\class{\varphi}_S^\alpha$.
				\end{description}
			\end{multicols}
		\end{quote}
		It follows that for every $\varphi::S$, $\class{\varphi}_S^\alpha$ is measurable.
	\end{defi}
	
	\begin{defi}\label{interpretationsofformulas}
		For each pair of numbers $p,q\in[0,1]$, the modal operator $[(p,q)]$,  allows us to express that for the interpretation of a formula $\varphi::S$ the upper probability is bigger or equal than $p$, and the lower probability is bigger or equal than $q$. We introduce for readability the following notation which may be found in \cite{dodersavic2015}. For an upper probability measure $g\in \Delta^*SX$, and its corresponding lower measure $\underline{g}(V)=1-g(V^c)$ for all $V\in \Sigma_{SX}$ we write:

		\begin{itemize}
			\item $\upper{\ge p}\varphi$ instead of $[(p,0)]\varphi$. Notice that $g\in\class{[(p,0)]\varphi}=\class{\upper{\ge p}\varphi}$ iff $g(\class{\varphi})\ge p$ and $1-g(\class{\varphi}^c)\ge 0$, which translates to $g(\class{\varphi})\ge p$ and $\underline{g}(\class{\varphi})\ge 0$. Since the second condition is always true, this is just $g(\class{\varphi})\ge p$.
		\end{itemize}
		Similarly, we write
		\begin{itemize}
			\item $\plower{\ge p}\varphi$ for $[(0,p)]\varphi$. Then $g\in\class{[(0,p)]\varphi}$ iff $g(\class{\varphi})\ge 0$ and $\underline{g}(\class{\varphi})\ge p$, i.e. $\underline{g}(\class{\varphi})\ge p$.
			\item $\upper{\le p}\varphi$ for $[(0,1-p)]\neg\varphi$, so $g\in\class{\upper{\le p}\varphi}$ iff $g(\class{\varphi})\le p$.
			\item $\plower{\le p}\varphi$ for $[(1-p,0)]\neg\varphi$, so $g\in\class{\upper{\le p}\varphi}$ iff $g(\class{\varphi}^c)\ge 1-p$, i.e. $\underline{g}(\class{\varphi})\le p$.
			\item $\upper{< p}\varphi$ for $\neg[(p,0)]\varphi$, so $g\in\class{\upper{< p}\varphi}$ iff $g(\class{\varphi})<p$.
			\item $\plower{<p}\varphi$ for $\neg[(0,p)]\varphi$, so $g\in\class{\plower{<p}\varphi}$ iff $\underline{g}(\class{\varphi})<p$.
			\item $\upper{>p}\varphi$ for $\neg[(0,1-p)]\neg\varphi$, so  $g\in\class{\upper{>p}\varphi}$ iff $g(\class{\varphi})>p$.
			\item $\plower{>p}\varphi$ for $\neg[(1-p,0)]\neg\varphi$, so $g\in\class{\plower{>p}\neg\varphi}$ iff $\underline{g}(\class{\varphi})>p$.
		\end{itemize}
	\end{defi}
	
	\begin{obs}\label{changeofnotation}
		Using the abbreviations of Definition \ref{interpretationsofformulas}, we replace the formulas with the modal operators $[p,q]$ with others that refer to bounds on the upper or lower probabilities.
		\begin{itemize}
			\item The formula $[(p,0)]\varphi$ can be written both as $\upper{\ge p}\varphi$ and as $\neg\upper{<p}\varphi$. Similarly, $[(0,q)]\varphi$ can be written as $\plower{\ge q}\varphi$ and as $\neg\plower{<q}\varphi$. These equivalences reflect that the upper and lower probabilities of an event are real numbers and therefore linearly ordered.
			\item In the same manner, $[(1-p,0)]\neg\varphi$ can be written as $\upper{\ge 1-p}\neg\varphi$ or as $\plower{\le p}\varphi$, while both $\plower{\ge 1-q}\neg\varphi$ and $\upper{\le q}\varphi$ stand for $[(0,1-q)]\neg\varphi$. Here we see explicitly the duality between upper and lower probabilities from Lemma \ref{upperprobproperties} \ref{upperpropduality}.
		\end{itemize}
		
		We will use any of the equivalent forms interchangeably in the proofs. 
	\end{obs}
	
	$T$-coalgebra homomorphisms preserve the semantics of the formulas:
	
	\begin{lem}\label{preservation}
		If $f:(X,\alpha)\to(Y,\delta)$ is a $T$-coalgebra morphism, and $S\in\ing(T)$, then for every formula $\varphi:S$, $(Sf)^{-1}\class{\varphi}^\delta_S=\class{\varphi}^\alpha_S$. In particular, for $\varphi:Id$, $x\in\class{\varphi}^\alpha_S$ iff $f(x)\in\class{\varphi}^\delta_S$.
	\end{lem}
	\begin{proof}
		The proof goes by recursion on the ingredients and formulas. We present here two steps: 
		
		By Lemma \ref{losbetasmedibles}, $(\Delta^*Sf)^{-1}\class{[(p,q)]\varphi}^{\delta}_{\Delta^*S}=(\Delta^*Sf)^{-1}\beta^{p,q}\class{\varphi}^{\delta}_{S}=\beta^{p,q}(Sf)^{-1}\class{\varphi}^{\delta}_{S}=\beta^{p,q}\class{\varphi}^{\alpha}_{S}=\class{[(p,q)]\varphi}^{\alpha}_{\Delta^*S}$.
		
		By the definition of coalgebra, we have that $(Id\ f)^{-1}\class{[\sig]\varphi}^{\delta}_{Id}=f^{-1}\class{[\sig]\varphi}^{\delta}_{Id}=f^{-1}\delta^{-1}\class{\varphi}^{\delta}_{T}=(\delta f)^{-1}\class{\varphi}^{\delta}_{T}
		=(Tf\alpha)^{-1}\class{\varphi}^{\delta}_{T}=\alpha^{-1}(Tf)^{-1}\class{\varphi}^{\delta}_{T}=\alpha^{-1}\class{\varphi}^{\alpha}_{T} =\class{[\sig]\varphi}^{\alpha}_{Id}$.
	\end{proof}
	
	We will write $\class{\varphi}$ instead of $\class{\varphi}_S^\alpha$, when it is clear from the context.
	
	\begin{defi}[\bf{Satisfaction}]\label{defsatisfaction}
		
		Given a $T$-coalgebra $(X,\alpha)$, for each $S\in\ing T$, we say that an element $x\in SX$ \emph{satisfies} the formula $\varphi$, and we write $\alpha,x\sat_S\varphi$, if $x\in\class{\varphi}_S^\alpha$. This gives
		
		\[
		\alpha,x\not\sat_S\bot_S,
		\ \ \ \alpha,x\sat_{M}A\text{ iff }x\in A,
		\]
		\[
		\alpha,x\sat_S\varphi_1\to\varphi_2\text{ iff }\alpha,x\sat_S\varphi_1\text{ implies }\alpha,x\sat_S\varphi_2,
		\]
		\[
		\alpha,x\sat_{S_1\times S_2}[\pr_j]\varphi\text{ iff }\alpha,\pi_j(x)\sat_{S_j}\varphi,
		\]
		\[
		\alpha,x\sat_{S_1+S_2}[\inc_j]\varphi\text{ iff }x=in_j(y)\text{ implies }\alpha,y\sat_{S_j}\varphi,
		\]
		\[
		\alpha,g\sat_{\Delta^*S}\upper{\ge p}\varphi\text{ iff }g(\class{\varphi}_S^\alpha)\ge p,\ \ \ \ \ \ 
		\alpha,g\sat_{\Delta^*S}\plower{\ge q}\varphi\text{ iff }\underline{g}(\class{\varphi}_S^\alpha)\ge q,
		\]
		\[
		\alpha,g\sat_{\Delta^*S}[(p,q)]\varphi\text{ iff }\alpha,g\sat_{\Delta^*S}\upper{\ge p}\varphi\text{ and }\alpha,g\sat_{\Delta^*S}\plower{\ge q}\varphi.
		\]
	\end{defi}
	
	The semantics of the formulas $[(p,0)]\varphi$ and $[(0,q)]\varphi$ are a natural generalization of satisfaction for modal probabilistic formulas, similar to the one introduced by Larsen and Skou \cite{LarsenSkou91} in the case of discrete state spaces.
	
	\begin{obs}\label{relationskripke}
		For $\kappa\ne(p,q)$, the semantics of $[\kappa]\varphi:S$ can be read as the usual Kripke semantics using the elements of $SX$ as worlds in a frame and an accessibility relation $R_\kappa\subseteq SX\times S'X$. For example, if $S=S_1\times S_2$, and $\kappa=\pi_1$, then $R_{\pi_1}\subseteq(S_1\times S_2)X\times S_1X$, and for $x\in S_1X$, $xR_{\pi_1}y$ iff $y=\pi_1(x)$.
	\end{obs}
	
	Some notation required for the rest of the paper: if $\Gamma\subseteq\mathsf{Form}_S$, then $\alpha,x\sat_S\Gamma$ indicates that $\alpha,x\sat_S\varphi$, for all $\varphi\in\Gamma$. Furthermore, $\Gamma\sat_S^\alpha\varphi$ means that for all $x\in SX$ such that $\alpha,x\sat_S\Gamma$, we have that $\alpha,x\sat_S\varphi$. Finally, the notation $\Gamma\sat_S\varphi$ is short for $\Gamma\sat_S^\alpha\varphi$ for all $T$-coalgebras $\alpha$. We will also use the description sets
	$des_S^\alpha(x)=\set{\varphi:S\st\alpha,x\sat_S\varphi}$ to prove completeness for the logic defined later in this work.
	
	\begin{exa}
		The interval-valued finite Markov chains presented in \cite{kozine02interval} can be regarded as coalgebras $c:X\to \Delta^*X$, in which the map $c$ assign to each state in the set $X$ an upper probability on all the states. 
	\end{exa}
	
	\begin{exa}
		To get a better grasp of the ideas used in this work, consider the functor $T=\Delta^*(Id\times M)$, where $M=\{a,b,c\}$ is an uncertainty space with the algebra $\Sigma_M=\{\emptyset,\{a\},\{b,c\}, M\}$. We take the uncertainty space $X=(X,\Sigma_X)$ with $X=\{x,y,z,t\}$ and $\Sigma_X=\{\emptyset,\{x,y\},\{z,t\},X\}$. Then  $TX$ is the space $\Delta^*(X\times M)$, with the algebra generated by the family $\{U\times V| U\in\Sigma_X,V\in\Sigma_M\}$.
		
		Consider the probabilities defined for the generators of the algebra in the following table:		
		\begin{center}
			\begin{tabular}{c||c|c|c|c}
				$\ $ & $\mu_1$ & $\mu_2$ & $\mu_3$ & $\mu_4$\\ \hline\hline
				$\set{x,y}\times\set{a}$ & $0.2$ & $0.4$ & $0.25$ & $0$\\ \hline
				$\set{z,t}\times\set{a}$ & $0.1$ & $0.2$ & $0$ & $0.3$\\ \hline
				$\set{x,y}\times\set{b,c}$ & $0$ & $0.2$ & $0.25$ & $0.4$\\ \hline
				$\set{z,t}\times\set{b,c}$ & $0.7$ & $0.2$ & $0.5$ & $0.3$\\ 
			\end{tabular}
		\end{center}
		
		Taking $\mathcal{P}_1=\{\mu_1,\mu_2\}$, $\mathcal{P}_2=\{\mu_3,\mu_4\}$, we define $\alpha(x)=\alpha(y)=\mathcal{P}^*_1$ and $\alpha(z)=\alpha(t)=\mathcal{P}^*_2$. It is easy to check that $\alpha:X\to TX$ is a measurable function, so $(X,\alpha)$ is a $T$-coalgebra. We have that $\ing T=\{Id,M,Id\times M,T\}$.
		
		The associated multigraph of ingredients is the following:
		
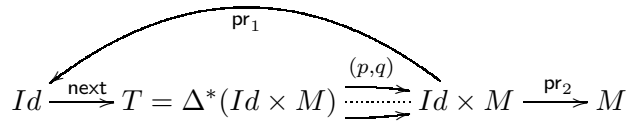
\[
		{
			\xymatrix{Id\ar[r]^(.3){\sig}&T=\Delta^*(Id\times M)\ar@/^/[r]^<(.2){(p,q)}\ar@/_/[r]\ar@{.}[r]
				&Id\times M\ar[r]^<(.4){\pr_2}\ar@/_3pc/[ll]^{\pr_1}&M}
		}
		\]
		Some formulas of different sorts are given in the table below:
		\begin{center}
			\begin{tabular}{c|l} 
				$M$ & $\bot_M$,  $\{a\}$, $\{b\}$, $\{c\}$,\\
				& $\{b,c\}$,  $M$,\\
				& $\neg\{b,c\}$, $\neg\{a\}$, $\{b\}\lor\{b,c\}$\\ \hline
				$Id$ & $\bot_{Id}$, $\top_{Id}=\bot_{Id}\to\bot_{Id}$,\\
				& $[\sig][(0.5,0.4)][\pr_2]\neg\{b,c\}$,\\
				& $[\sig][(0,1)][\pr_1]\bot_{Id}\lor[\sig][(1,0)][\pr_2]\{b,c\}$\\ \hline
				$Id\times M$ & $\bot_{Id\times M}$, $\top_{Id\times M}$,\\
				& $[\pr_1]\top_{Id}$,$[\pr_2]\neg\{b,c\}$, $[\pr_2]\top_M$,\\
				& $[\pr_1][\sig][(0.5,0.4)][\pr_2]\neg\{b,c\}$\\ \hline 
				$T=\Delta^*(Id\times M)$ & $\bot_T$, $\top_T$, \\
				& $[(0.5,0.4)][\pr_2]\neg\{b,c\}$,\\
				& $\upper{\ge 0.5}[\pr_2]\neg\{b,c\}\land \plower{\ge 0.4}[\pr_2]\neg\{b,c\}$
			\end{tabular}
		\end{center}
		
		Notice that $\neg\{a\}::M$, but since $\{b\}$ is not measurable, $\{b\}\lor\{b,c\}\not::M$, even tough they have the same interpretation.
		
		The formula $\varphi=[\sig][(0.5,0.4)][\pr_2]\neg\{b,c\}$ is built by going through a path in the multigraph from $Id$ to $M$. Notice that by going through a cycle, modal formulas of arbitrary length can be formed. 
		
		What is the interpretation in the coalgebra $(X,\alpha)$ of $\varphi$?. To see this, we write the interpretation of each of its subformulas as follows:
		\begin{itemize}
			\item $\class{\neg\{b,c\}}^\alpha_M=\{b,c\}^c=\{a\}$.
			\item $\class{[\pr_2]\neg\{b,c\}}^\alpha_{Id\times M}=\pi_2^{-1}(\{a\})=X\times\{a\}$.
			\item $\class{[(0.5,0.4)][\pr_2]\neg\{b,c\}}^\alpha_T=$ $\{g\in\Delta^*(X\times M)| g(X\times\{a\})\ge0.5, 1-g((X\times\{a\})^c)\ge0.4\}=$ $\{g\in\Delta^*(X\times M)| g(X\times\{a\})\ge0.5, g(X\times\{b,c\})\le0.6\}$.
			\item $\class{[\sig][(0.5,0.4)][\pr_2]\neg\{b,c\}}^\alpha_{Id}=\alpha^{-1}(\{g\in\Delta^*(X\times M)| g(X\times\{a\})\ge 0.5,g(X\times\{b,c\})\le 0.6\})=\alpha^{-1}(\{\mathcal{P}^*_1\})=\{x,y\}$.
		\end{itemize}
		
		From the above, $[(0.5,0.4)][\pr_2]\neg\{b,c\}\in des_{T}^\alpha(\mathcal{P}^*_1)$ and $\varphi\in des_{Id}^\alpha(x)$.
		
		The formula $[\sig](\upper{\ge 0.5}[\pr_2]\neg\{b,c\}\land\plower{\ge 0.4}[\pr_2]\neg\{b,c\})$, has the same semantics.
	\end{exa}
	
	\subsection{Axioms and deduction systems} \label{axiomsanddeductivesystems}
	
	\begin{defi}\label{axiomsupprob}
		We now define for each $S\in\ing T$, the set of axioms $Ax_{S}\subseteq\mathsf{Form}_S$, consisting of the formulas:
		\begin{enumerate}[\normalfont(1)]
			\item All the Boolean tautologies $\varphi:S$,\label{axtaut}
			\item for $S=M$, $A:M$ and $c\in M$,\label{axconstant}
			\begin{enumerate}[\normalfont(a)]
				\item $\{c\}\to A$ if $c\in A$,\label{axconstant1}
				\item $\{c\}\to\neg A$ if $c\notin A$,\label{axconstant2}
			\end{enumerate}
			\item for $S=S_1\times S_2$, $j\in\{1,2\}$ and $\varphi:S_j$,\label{axprod}
			\begin{enumerate}[\normalfont(a)]
				\item $\neg[\pr_j]\varphi\to[\pr_j]\neg\varphi$,\label{axprod1}
				\item $\neg[\pr_j]\bot_{S_j}$,\label{axprod2}
			\end{enumerate}
			\item for $S=S_1+S_2$,\label{axcoprod}
			\begin{enumerate}[\normalfont(a)]
				\item $\neg[\inc_j]\varphi\to[\inc_j]\neg\varphi$,\label{axcoprod1}
				\item $\neg[\inc_1]\bot_{S_1}\leftrightarrow[\inc_2]\bot_{S_2}$,\label{axcoprod2}
			\end{enumerate}
			\item for $S=Id$, and $\varphi:T$,\label{axident}
			\begin{enumerate}[\normalfont(a)]
				\item $\neg[\sig]\varphi\to[\sig]\neg\varphi$,\label{axident1}
				\item $\neg[\sig]\bot_T$,\label{axident2}
			\end{enumerate}
			\item for $S=\Delta^*S'$,\label{axupprob}
			\begin{enumerate}[\normalfont(a)]
				\item $\upper{\le p}\varphi\to\upper{<q}\varphi$ for $p<q$,\label{axupprob1}
				\item $\upper{<p}\varphi\to\upper{\le p}\varphi$,\label{axupprob2}
				\item $\plower{\ge1}(\varphi\to\psi)\to(\upper{\ge p}\varphi\to\upper{\ge p}\psi)$,\label{axupprob5}
				\item $[(p,q)]\varphi\leftrightarrow(\upper{\ge p}\varphi\land\plower{\ge q}\varphi)$.\label{axupprob6}
			\end{enumerate}
		\end{enumerate}
	\end{defi}
	
	Axiom \ref{axtaut} makes our logic classic at every ingredient. Axiom \ref{axconstant} is necessary to create a copy of each uncertainty space that appears as an ingredient of $T$ when we build the \emph{canonical coalgebra} out of sets of formulas in Definition \ref{defcanonicalcoalgebra}.
	
	Axioms \ref{axprod}\ref{axprod1} to \ref{axident}\ref{axident1} correspond to the fact that the accessibility relations $R_\kappa$ described in Observation \ref{relationskripke} are functional, that is, for each $x$ such that $\alpha,x\sat_S[\kappa]\varphi$, there is at most one $y$ such that $xR_\kappa y$, and $\alpha,y\sat_{S'}\varphi$, while the part (b) of those axioms say that that the relations, except in the case of the coproduct, are total. For the coproduct, axiom \ref{axcoprod}\ref{axcoprod2} indicates that the relation $R_{in_i}$ is the identity on the $i$-th summand of the coproduct.
	
	Axiom \ref{axupprob} characterizes the upper (and lower) probabilities: items \ref{axupprob1} and \ref{axupprob2} are adapted from the proposed by Savić, Doder, Ognjanović in \cite{dodersavic2015}. 
	Axiom \ref{axupprob}\ref{axupprob5} implies that if $\varphi$ and $\psi$ are equivalent formulas, they will have the same upper probability. Finally, axiom \ref{axupprob}\ref{axupprob6} allows us to separate the formula $[p,q]\varphi$, which gives bounds for both the lower and upper probability of $\varphi$ simultaneously, as the conjunction of two formulas, each one focusing only on the lower or the upper probability separately. This will also be necessary to prove the completeness of the logic.
	
	\begin{obs}\label{passageofformulas}
		In Observation \ref{changeofnotation}, we saw some equivalent ways of writing formulas of the form $[(p,0)]\varphi$ or formulas of the form $[(0,q)]\psi$, but the only way we have of connecting these two kinds of formulas is through the axioms: by \ref{axupprob}\ref{axupprob1} if $p<q$, $[(0,1-q)]\neg\varphi\to\neg[(q,0)]\varphi$ and by \ref{axupprob}\ref{axupprob2} $\neg[(p,0)]\varphi\to [(0,1-p)]\neg\varphi$.
	\end{obs}
	
	\begin{thm}\label{axiomsarevalid}
		For each ingredient $S$ of $T$, all axioms of sort $S$ are valid in all $T$-coalgebras.
	\end{thm}
	
	\begin{proof}
		The validity of axioms \ref{axtaut}-\ref{axident} are a direct consequence of the semantics defined. Axiom \ref{axupprob}\ref{axupprob1} says that the upper and lower probability measures are greater or equal than 0, which is true by Lemma \ref{upperprobproperties}. Axioms \ref{axupprob}\ref{axupprob1} and \ref{axupprob}\ref{axupprob2} follow from the simple observations that $g(U)\le p$ implies $g(U)<q$ for every $p<q$, and $g(U)<p$ implies $g(U)\le p$, respectively.
		
		For axiom \ref{axupprob}\ref{axupprob5}, assume that (iv) $\alpha,g\sat_{\Delta^*S'}\plower{\ge1}(\varphi\to\psi)$ and (v) $\alpha,g\sat_{\Delta^*S'}\upper{\ge p}\varphi$. We want to show that $\alpha,g\sat_{\Delta^*S'}\upper{\ge p}\psi$, which is equivalent to show that $g(\class{\psi})\ge p$.  
		From (iv), $1-g(\class{\varphi\to\psi}^c)=1$, so $1-g(\class{\varphi\land\neg\psi})=1$, and this implies that (vi) $g(\class{\varphi\land\neg\psi})=0$. On the other hand, (v) is equivalent to $g(\class{\varphi})\ge p$. 
		Hence, using subadditivity and monotonicity, it follows that $p\le g(\class{\varphi})=g(\class{\varphi\land\psi}\cup\class{\varphi\land\neg\psi})\le g(\class{\varphi\land\psi})+g(\class{\varphi\land\neg\psi})$ $\underset{\text{(vi)}}{=}g(\class{\varphi\land\psi})\le g(\class{\psi})$. 
		Therefore, $\alpha,g\sat_{\Delta^*S'}\plower{\ge1}(\varphi\to\psi)\to(\upper{\ge p}\varphi\to\upper{\ge p}\psi)$.
		
		Axiom \ref{axupprob}\ref{axupprob6} is easy to verify from the semantics.
	\end{proof}
	
	\begin{defi}\label{defdedsystem}
		Adapting the definition by Goldblatt \cite{goldblattdeduction2010}, we say that a set of relations $\{\ded_S\subseteq$\linebreak[3]$\mathcal{P}(\mathsf{Form}_S)\times\mathsf{Form}_S|S\in\ing T\}$ is a $T$-deduction system if the following rules hold:
		\begin{itemize}
			\item \underline{Assumption rule:} If $\varphi\in\Gamma\cup Ax_{S}$, then $\Gamma\ded_S\varphi$.
			\item \underline{Modus ponens:}  $\{\varphi,\varphi\to\psi\}\ded_S\psi$.
			\item \underline{Cut rule:} If $\Gamma\ded_S\Lambda$ and $\Lambda\ded_S\varphi$, then $\Gamma\ded_S\varphi$.
			\item \underline{Deduction rule:} If $\Gamma\cup\{\varphi\}\ded_S\psi$, then $\Gamma\ded_S\varphi\to\psi$.
			\item \underline{Necessitation rule:} If there is an edge $\Delta^*S\overset{(0,1)}{\rightsquigarrow}S$, and $\ded_S\varphi$, then $\ded_{\Delta^*S}\plower{\ge1}\varphi$.
			\item \underline{Constant rule:} If $M\in\ing T$, then $\{\neg\{c\}\st c\in M\}\ded_{M}\bot_{M}$.
			\item \underline{Definite box rule:} For each $\kappa\ne(p,q)$, $S\overset{\kappa}{\rightsquigarrow}S'$, $\Gamma\ded_{S'}\psi$ implies $[\kappa]\Gamma\ded_S[\kappa]\psi$, where $[\kappa]\Gamma=\{\kappa\psi|\kappa\in\Gamma\}$.
			\item \underline{Archimedean rules:} $\{\upper{\ge q}\psi\st q<p\}\ded_{\Delta^*S}\upper{\ge p}\psi$. Also, $\{\plower{\ge q}\psi\st q<p\}\ded_{\Delta^*S}\plower{\ge p}\psi$.
			\item \underline{Cover rules:} 
			\begin{enumerate}[\normalfont(1)]
				\item If $n\ge 1$, $\ded_{S}\varphi\to\bigvee_{I\subseteq J_m,\ \vabs{I}=n+k}\bigwedge_{i\in I}\varphi_i$, $\ded_{S}\bigvee_{I\subseteq J_m,\ \vabs{I}=k}\bigwedge_{i\in I}\varphi_i$, $p_1,\ldots,p_m\in [0,1]$, and $p=0\lor(1\land\frac{\sum_{i=1}^mp_i-k}{n})$, then
				\[\ded_{\Delta^*S}(\upper{\le p_1}\varphi_1\land\cdots\land\upper{\le p_m}\varphi_m)\to\upper{\le p}\varphi.\] \label{nkcoverrule1}
				\item If $\ded_S\bigvee_{I\subseteq J_m, \vabs{I}=k}\bigwedge_{i\in I}\varphi_i$, and $\sum_{i=1}^mp_i<k$ (with $ p_i\in [0,1]$), then \[\ded_{\Delta^*S}\neg(\upper{\le p_1}\varphi_1\land\cdots\land\upper{\le p_m}\varphi_m).\]\label{nkcoverrule2}
			\end{enumerate}
		\end{itemize}
	\end{defi}
	\begin{obs}  This definition differs from the one given by Goldblatt in that he  includes a deduction rule to account for the countable additivity property of probability measures, and we added a rule of necessitation for the upper probability modal operators. The cover rules correspond to the condition \ref{30032024-3}. The second part is needed to handle the case in which $n=0$ or the set corresponding to the formula $\varphi$ is assigned measure $0$.
		
		By its nature, the deduction system presented here is not finitary and doesn't have the compactness property.
	\end{obs}
	Some of the consequences of this definition are collected in the next lemma.
	
	\begin{lemC}[\cite{goldblattmodality93}]\label{propsistdeduct}\ 
		\begin{enumerate}[\normalfont(1)]
			\item \underline{Monotonicity:} If $\Gamma\ded_S\varphi$ and $\Gamma\subseteq\Theta$, then $\Theta\ded_S\varphi$.\label{modalmonotonicity}
			\item \underline{Detachment:} If $\Gamma\ded_S\varphi$ and $\Gamma\ded_S\varphi\to\psi$, then $\Gamma\ded_S\psi$.\label{modaldetachment}
			\item \underline{Converse deduction:} $\Gamma\ded_S\varphi\to\psi$ implies $\Gamma\cup\{\varphi\}\ded_S\psi$.\label{modalconversededuction}
			\item If $\Gamma\ded_S\varphi$ and $\Gamma\cup\{\varphi\}\ded_S\bot_S$, then $\Gamma\ded_S\bot_S$.\label{modalsdprop1}
			\item $\Gamma\cup\{\neg\varphi\}\ded_S\bot_S$ iff $\Gamma\ded_S\varphi$.\label{modalsdprop2}
			\item \underline{Implication rule:} $\Gamma\ded_S\varphi$ implies $\psi\to\Gamma\ded_S\psi\to\varphi$.\label{modalimplicationrule}
		\end{enumerate}
	\end{lemC}
	
	\begin{proof} \ref{modalmonotonicity} Notice that if $\psi\in\Gamma$, then the assumption rule gives $\Gamma\ded_S\psi$. Since $\Gamma\subseteq\Theta$, it follows that $\Theta\ded_S\Gamma$. Using the cut rule, we have that $\Theta\ded_S\varphi$.
		
		\ref{modaldetachment} From the hypothesis  $\Gamma\ded_S\{\varphi,\varphi\to\psi\}$, and by modus ponens, $\{\varphi,\varphi\to\psi\}\ded_S\psi$, so by the cut rule it follows that $\Gamma\ded_S\psi$.
		
		\ref{modalconversededuction} If $\Gamma\ded_S\varphi\to\psi$, by monotonicity we have that $\Gamma\cup\{\varphi\}\ded_S\varphi\to\psi$. On the other hand, by the assumption rule, it follows that $\Gamma\cup\{\varphi\}\ded_S\varphi$. Now, since $\{\varphi,\varphi\to\psi\}\ded_S\psi$ (modus ponens), the cut rule implies that $\Gamma\cup\{\varphi\}\ded_S\psi$.
		
		\ref{modalsdprop1} If $\Gamma\ded_S\varphi$ and $\Gamma\cup\{\varphi\}\ded_S\bot_S$, then $\Gamma\ded_S\varphi\to\bot_S$ by the deduction rule. Therefore, $\Gamma\ded_S\{\varphi,\varphi\to\bot_S\}$, and since $\{\varphi,\varphi\to\bot_S\}\ded_S\bot_S$, the cut rule yields $\Gamma\ded_S\bot_S$.
		
		\ref{modalsdprop2} This follows easily from the tautology $\neg\neg\varphi\leftrightarrow\varphi$, deduction and converse deduction rules.
		
		\ref{modalimplicationrule} By modus ponens $\{\psi\to\zeta,\psi\}\ded_S\zeta$, for all $\zeta\in\Gamma$, and by monotonicity we have that $(\psi\to\Gamma)\cup\{\psi\}\ded_S\Gamma$. By hypothesis, $\Gamma\ded_S\varphi$, so the cut rule ensures that $(\psi\to\Gamma)\cup\{\psi\}\ded_S\varphi$. Hence, the deduction rule yields $\psi\to\Gamma\ded_S\psi\to\varphi$.
	\end{proof}
	
	\begin{obs}
		The assumption rule and detachment allow us to use boolean reasoning naturally. For example, from the assumption rule, the axiom $(\varphi\to\psi)\leftrightarrow(\neg\psi\to\neg\varphi)$ implies that $\ded_S(\varphi\to\psi)\leftrightarrow(\neg\psi\to\neg\varphi)$, and if $\ded_S\varphi\to\psi$, we can conclude $\ded_S\neg\psi\to\neg\varphi$. Therefore, we will sometimes omit some deduction steps just appeal to ``boolean reasoning''.
	\end{obs}
	
	\begin{defi}\label{defsoundness}
		A $T$-deduction system is \emph{sound} if $\Gamma\ded_S\varphi$ implies $\Gamma\sat_S\varphi$ for all $S\in\ing T$.
	\end{defi}
	
	\begin{thm}\label{conseqT} The inference rules of $T$-deduction systems preserve validity.
	\end{thm}
	\begin{proof}
		We must prove that for any $T$-coalgebra $(X,\alpha)$, the rules of Definition \ref{defdedsystem} preserve validity. By Theorem \ref{axiomsarevalid}, the assumption rule is straightforward. The modus ponens, cut, deduction, constant and definite box rules follow from Definition \ref{defsatisfaction}.
		
		We prove now that cover rule \ref{nkcoverrule1} is valid. If $\sat_{\Delta^*S}^\alpha\varphi\to\bigvee_{I\subseteq J_m, \vabs{I}=n+k}\bigwedge_{i\in I}\varphi_i$, then $\class{\varphi}\subseteq\bigcup_{I\subseteq J_m, |I|=n+k}\bigcap_{i\in I}\class{\varphi_i}$, so $\class{\varphi}$ is covered $n+k$ times by the sequence $\class{\varphi_1},\ldots,\class{\varphi_m}$ (Definition \ref{nkcover}), and analogously if $\sat_{\Delta^*S}^\alpha\bigvee_{I\subseteq J_m, \vabs{I}=k}\bigwedge_{i\in I}\varphi_i$, then $SX=\class{\top_{S}}$ is covered $k$ times by $\class{\varphi_1},\ldots,\class{\varphi_m}$.
		
		To check that $(\upper{\le p_1}\varphi_1\land\cdots\land\upper{\le p_m}\varphi_m)\to\upper{\le p}\varphi$ is valid in a coalgebra $(X,\alpha)$, we need to prove that, for any $g\in\Delta^*SX$, if we assume that (i) $\alpha,g\sat_{\Delta^*S}\upper{\le p_1}\varphi_1\land\cdots\land\upper{\le p_m}\varphi_m$, then (ii) $\alpha,g\sat_{\Delta^*S}\upper{\le p}\varphi$. From (i), it follows that $\alpha,g\sat_{\Delta^*S}\upper{\le p_i}\varphi_i$ for $i\in J_m$, so  $g(\class{\varphi_i})\le p_i$ due to definitions \ref{defsatisfaction} and \ref{interpretationsofformulas}. Therefore
		\[\sum_{i=1}^mg(\class{\varphi_i})\le\sum_{i=1}^mp_i,\]
		and using \ref{30032024-3},
		\[k+ng(\class{\varphi})\le\sum_{i=1}^mg(\class{\varphi_i})\le\sum_{i=1}^mp_i,\]
		so 
		\[g(\class{\varphi})\le0\lor\left(1\land\frac{\sum_{i=1}^mp_i-k}{n}\right)=p,\]
		which is equivalent to $\alpha,g\sat_{\Delta^*S}\upper{\le p}\varphi$.
		
		To see the validity of the cover rule \ref{nkcoverrule2}, assume that $\sat_{\Delta^*S}^\alpha\bigvee_{I\subseteq J_m, \vabs{I}=k}\bigwedge_{i\in I}\varphi_i$ and (iii) $\sum_{i=1}^mp_i<k$. If $\alpha,g\sat_{\Delta^*{S'}}(\upper{\le p_1}\varphi_1\land\cdots\land\upper{\le p_m}\varphi_m)$, then for all $i\in J_m$ we have that $g(\class{\varphi_i})\le p_i$, so $\sum_{i=1}^mg(\class{\varphi_i})\le\sum_{i=1}^mp_i$. Now, consider the sequence $\class{\varphi_1},\ldots,\class{\varphi_m}$. From the hypothesis, it is a $(0,k)$-cover of $(\class{\top_{S}},\class{\top_{S}})$. So, by \ref{30032024-3} it follows that $k\le\sum_{i=1}^mg(\class{\varphi_i})\le\sum_{i=1}^mp_i$, which contradicts (iii). Therefore, $\alpha,g\sat_{\Delta^*S}\neg(\upper{\le p_1}\varphi_1\land\cdots\land\upper{\le p_m}\varphi_m)$.
		
		For the Necessitation rule, suppose that $\sat_S^\alpha\varphi$. We need to check that $\sat_{\Delta^*S}^\alpha\plower{\ge1}\varphi$.	The hyphotesis means that $\class{\varphi}=SX$, so for every $g\in\Delta^*S$, $\underline{g}(\class{\varphi})=\underline{g}(SX)=1-g((SX)^c)=1$. This proves that $\alpha,g\sat_{\Delta^*S}\plower{\ge1}\varphi$ for all $g\in\Delta^*S$ and therefore $\sat_S^\alpha\plower{\ge1}\varphi$.
		
		The Archimedean rule $\{\upper{\ge q}\varphi\st q<p\}\sat_{\Delta^*S}^\alpha\upper{\ge p}\varphi$ is equivalent to the fact that for every $g\in\Delta^*SX$, if $g(\class{\varphi})\ge q$ for all $q<p$, then $g(\class{\varphi})\ge p$, but this is true by the archimedean property on the real numbers. Analogously the other Archimedean rule follows.
	\end{proof}
	
	\begin{cor}
		For each $T$-coalgebra $(X,\alpha)$, the family of semantic consequence relations $Cons_T^\alpha=\{\sat_S^\alpha\st S\in\ing T\}$, and $Cons_T=\{\sat_S\st S\in\ing T\}$ are $T$-deduction systems.
	\end{cor}
	
	As a direct consequence of Theorems \ref{axiomsarevalid} and \ref{conseqT}, we have:
	
	\begin{thm}\label{soundness}
		$T$-deduction systems are sound with respect to the coalgebraic semantics from Definition \ref{defsemantics}.
	\end{thm}
	
	We will prove now some more consequences of the definitions:
	
	\begin{lem}\label{dedproplema} For any formulas $\varphi,\psi::S$:
		\begin{enumerate}[\normalfont(1)]
			\item $\ded_{\Delta^*S}\upper{\ge 0}\varphi$ and $\ded_{\Delta^*S}\plower{\ge 0}\varphi$.\label{prop1}
			\item $\ded_{\Delta^*S}\upper{\le p}\varphi\to\plower{\le p}\varphi$.\label{prop2}
			\item $\ded_{\Delta^*S}\plower{\ge p}\varphi\to\upper{\ge p}\varphi$.\label{prop2bis}
			\item If $p>0$, then $\ded_{\Delta^* S}\upper{<p}\bot_{S}$, and $\ded_{\Delta^*S}\plower{<p}\bot_{S}$.\label{prop3}
			\item If $\kappa\ne(p,q)$, $\ded_{S'}\psi$ implies $\ded_S[\kappa]\psi$.\label{prop4}
			\item If $\kappa\ne(p,q)$, $\ded_S[\kappa](\varphi\to\psi)\to([\kappa]\varphi\to[\kappa]\psi)$.\label{prop5}
			\item If $\kappa\ne(p,q)$, $\ded_{S'}\varphi\to\psi$ implies $\ded_S[\kappa]\varphi\to[\kappa]\psi$, and $\ded_{S'}\varphi\leftrightarrow\psi$ implies $\ded_S[\kappa]\varphi\leftrightarrow[\kappa]\psi$.\label{prop6}
			\item If $\ded_S\varphi\leftrightarrow\psi$, then $\ded_{\Delta^*S}\upper{\ge p}\varphi\leftrightarrow\upper{\ge p}\psi$.\label{prop5bis}
			\item $\ded_{\Delta^*S}\upper{\ge q}\varphi\to\upper{\ge p}\varphi$ if $q>p$.\label{prop7}
			\item $\ded_{\Delta^*S}\plower{\ge q}\varphi\to\plower{\ge p}\varphi$ if $q>p$.\label{prop8}
			\item If $\ded_S\varphi$, then $\ded_{\Delta^*S}\plower{\ge q}\varphi$ and $\ded_{\Delta^*S}\upper{\ge p}\varphi$.\label{prop9}
			\item If $\ded_S\varphi$, then  $\ded_{\Delta^*S}[(p,q)]\varphi$. In particular, $\ded_{\Delta^*S}[(p,q)]\top_S$.\label{prop10}
		\end{enumerate}
	\end{lem}
	
	\begin{proof}
		\ref{prop1} By definition of the formulas, $\set{\upper{\ge q}\varphi\st q<0}=\emptyset$. Then, the Archimedean rule ensures that $\set{\upper{\ge q}\varphi\st q<0}\ded_{\Delta^*S}\upper{\ge 0}\varphi$, that is $\ded_{\Delta^*S}\upper{\ge0}\varphi$. Analogously, we can prove that $\ded_{\Delta^*S}\plower{\ge 0}\varphi$.
		
		\ref{prop2} If $p=1$, we have $\ded_{\Delta^*S}\upper{\le1}\varphi\to\plower{\le1}\varphi$ which is equivalent to (see Observation \ref{changeofnotation}) $\ded_{\Delta^*S}\plower{\ge1-1}\neg\varphi\to\upper{\ge1-1}\neg\varphi$, that is $\ded_{\Delta^*S}\plower{\ge0}\neg\varphi\to\upper{\ge0}\neg\varphi$. But both $\upper{\ge0}\neg\varphi$ and $\plower{\ge0}\neg\varphi$ are abbreviations of $[(0,0)]\neg\varphi$, so the implication holds.
		
		Suppose that $p\ne1$. Then, we consider the tautology $\varphi\lor\neg\varphi=\bigvee_{I\subseteq\{1,2\},|I|=1}\bigwedge_{i\in I}\varphi_i$ (with $\varphi_1=\varphi$, $\varphi_2=\neg\varphi$), so by the cover rule \ref{nkcoverrule2} we have for $p+q<1$,
		\[\ded_{\Delta^*S}\neg(\upper{\le p}\varphi\land\upper{\le q}\neg\varphi),\]
		which is equivalent to
		\begin{equation}\label{03102021-1}
			\ded_{\Delta^*S}\upper{\le p}\varphi\to\neg\upper{\le q}\neg\varphi.
		\end{equation}
		By axiom \ref{axupprob}\ref{axupprob2} we have that $\ded_{\Delta^*S}\upper{<q}\neg\varphi\to\upper{\le q}\neg\varphi$, and by contraposition,
		\begin{equation}\label{03102021-2}
			\ded_{\Delta^*S}\neg\upper{\le q}\neg\varphi\to\neg\upper{<q}\neg\varphi.
		\end{equation}
		Then by \ref{03102021-1} and \ref{03102021-2}, for $p+q<1$:
		\begin{equation}\label{07102021-1}
			\ded_{\Delta^*S}\upper{\le p}\varphi\to\neg\upper{<q}\neg\varphi.
		\end{equation}
		By Observation \ref{changeofnotation}, \ref{07102021-1} is equivalent to
		\begin{equation}\label{26032022-1}
			\ded_{\Delta^*S}\upper{\le p}\varphi\to\upper{\ge q}\neg\varphi \ \ \text{for }q<1-p.
		\end{equation}
		By \ref{26032022-1} and the deduction rule, it follows that:
		\begin{equation}\label{15102021-1}
			\{\upper{\le p}\varphi\}\ded_{\Delta^*S}\upper{\ge q}\neg\varphi\ \ \text{for }q<1-p.
		\end{equation}
		
		Consider now the set of formulas  $\Gamma=\{\upper{\ge q}\neg\varphi\}_{q<1-p}$. Then, \ref{15102021-1} implies that $\upper{\le p}\varphi\ded_{\Delta^*S}\Gamma$. On the other hand, using that $q<1-p$, by the archimedean rule we obtain $\Gamma\ded_{\Delta^*S}\upper{\ge 1-p}\neg\varphi$. By the cut rule, we deduce that $\upper{\le p}\varphi\ded_{\Delta^*S}\upper{\ge 1-p}\neg\varphi$, so $\ded_{\Delta^*S}\upper{\le p}\varphi\to\upper{\ge 1-p}\neg\varphi$, which is equivalent by Observation \ref{changeofnotation} to $\ded_{\Delta^*S}\upper{\le p}\varphi\to\plower{\le p}\varphi$.
		
		\ref{prop2bis} By \ref{prop2}, we have that $\ded_{\Delta^*S}\upper{\le p}\varphi\to\plower{\le p}\varphi$, which is equivalent to $\ded_{\Delta^*S}\plower{\ge 1-p}\neg\varphi\to\upper{\ge1-p}\neg\varphi$. Replacing $p$ and $\varphi$ by $1-p$ and $\neg\varphi$ respectively, we obtain that $\ded_{\Delta^*S}\plower{\ge p}\varphi\to\upper{\ge p}\varphi$.
		
		\ref{prop3} Since $p>0$, by axiom \ref{axupprob}\ref{axupprob1}, we have that $\ded_{\Delta^*S}\upper{\le0}\bot_S\to\upper{<p}\bot_S$, and due to Observation \ref{changeofnotation}, $\ded_{\Delta^*S}\plower{\ge1-0}\neg\bot_S\to\upper{<p}\bot_S$, that is $\ded_{\Delta^*S}\plower{\ge1}\top_S\to\upper{<p}\bot_S$. On the other hand, using the Necessitation rule we obtain $\ded_{\Delta^*S}\plower{\ge1}\top_S$, and by Detachment, it follows that  $\ded_{\Delta^*S}\upper{<p}\bot_S$. By (3) and contraposition, we have that $\ded_{\Delta^*S}\neg\upper{\ge p}\bot_S\to\neg\plower{\ge p}\bot_S$, so $\ded_{\Delta^*S}\upper{<p}\bot_S\to\plower{<p}\bot_S$, and using Detachment we get  $\ded_{\Delta^*S}\plower{<p}\bot_S$.
		
		Items \ref{prop4}, \ref{prop5} and \ref{prop6} are proved in \cite{goldblattdeduction2010}, Lemma 4.6.
		
		\ref{prop5bis} Follows immediately from axiom \ref{axupprob}\ref{axupprob5}, using Necessitation and Detachment.
		
		\ref{prop7} If $q>p$, using axiom \ref{axupprob}\ref{axupprob1}, and contrapositive, we have that (i) $\ded_{\Delta^*S}\neg\upper{<q}\varphi\to\neg\upper{\le p}\varphi$. By axiom \ref{axupprob}\ref{axupprob2} and contrapositive it follows that (ii) $\ded_{\Delta^*S}\neg\upper{\le p}\varphi\to\neg\upper{<p}\varphi$. From (i) and (ii), we conclude that $\ded_{\Delta^*S}\neg\upper{<q}\varphi\to\neg\upper{<p}\varphi$, which by Observation \ref{changeofnotation} is equivalent to $\ded_{\Delta^*S}\upper{\ge q}\varphi\to\upper{\ge p}\varphi$.
		
		\ref{prop8} If $q>p$, then $1-q<1-p$, so using axiom \ref{axupprob}\ref{axupprob1} we have that (i) $\ded_{\Delta^*S}\upper{\le 1-q}\neg\varphi\to\upper{<1-p}\neg\varphi$. From axiom \ref{axupprob}\ref{axupprob2}, it follows that (ii) $\ded_{\Delta^*S}\upper{<1-p}\neg\varphi\to\upper{\le 1-p}\neg\varphi$. From (i) and (ii) we conclude that $\ded_{\Delta^*S}\upper{\le 1-q}\neg\varphi\to\upper{\le 1-p}\neg\varphi$, and by Observation \ref{changeofnotation} this is equivalent to $\ded_{\Delta^*S}\plower{\ge q}\varphi\to\plower{\ge p}\varphi$.
		
		\ref{prop9} The Necessitation rule and the hypothesis imply that $\ded_{\Delta^*S}\plower{\ge1}\varphi$. By item \ref{prop8}, it follows that (i) $\ded_{\Delta^*S}\plower{\ge q}\varphi$. Also, using item \ref{prop2bis} and Detachment, we deduce that $\ded_{\Delta^*S}\upper{\ge1}\varphi$, and by item \ref{prop7}, we have that $\ded_{\Delta^*S}\upper{\ge p}\varphi$. 
		
		\ref{prop10} Follows from \ref{prop9}, using axiom \ref{axupprob}\ref{axupprob6}. In particular, since $\top_S$ is a tautology, we deduce that $\ded_{\Delta^*S}[(p,q)]\top_S$.
	\end{proof}
	
	We are going to work with $T$-deduction systems that satisfy certain conditions, so we recall some standard notions for sets of formulas:
	
	\begin{defi}
		Let $\Gamma\subseteq\mathsf{Form}_S$, for $S\in\ing T$. If $Ax_S\subseteq\Gamma$,  and every time that $\varphi$ and $\varphi\to\psi\in\Gamma$ we have that $\psi\in\Gamma$ as well, then we say that $\Gamma$  is  an $S$-\emph{theory}.

		If for every $\varphi:S$, we have that $\varphi\in\Gamma$ iff $\neg\varphi\not\in\Gamma$, $\Gamma$ is \emph{negation complete}. Finally, if $\bot_S\notin\Gamma$, $\Gamma$ is $\bot$-free.
	\end{defi}
	
	\begin{lem}\label{proptheories}
		In every negation-complete $S$-theory $\Gamma$, we have that $\varphi\to\psi\in\Gamma$ iff $\varphi\in\Gamma$ implies $\psi\in\Gamma$. Moreover, if $\Gamma$ is a negation complete, $\bot$-free $S$-theory, then:
		\begin{multicols}{2}
			\begin{quote}
				$\neg\varphi$ iff $\varphi\notin\Gamma$;\\
				$\varphi\lor\psi\in\Gamma$ iff $\varphi\in\Gamma$ or $\psi\in\Gamma$;\\
				$\varphi\land\psi\in\Gamma$ iff $\varphi\in\Gamma$ and $\psi\in\Gamma$;\\
				$\varphi\leftrightarrow\psi\in\Gamma$ iff \emph{(}$\varphi\in\Gamma$ iff $\psi\in\Gamma$\emph{)}.
			\end{quote}
		\end{multicols}
		In particular, the sets $des_S^\alpha(x)$ are negation complete, $\bot$-free $S$-theories for every $T$-coalgebra $(X,\alpha)$.
	\end{lem}
	
	\begin{proof} Notice that all the tautologies from classical logic are in each $S$-theory by the assumption rule. Then, classical reasoning proves the first part of the lemma.		
		
		For the last assertion, consider the set $\Gamma^\alpha_S=\{\varphi:S\st\alpha\sat_S\varphi\}$. By Theorem \ref{soundness} all $S$-axioms are valid in all $T$-coalgebras, so $Ax_S\subseteq\Gamma^\alpha_S\subseteq des^\alpha_S(x)=\{\varphi:S\st\alpha,x\sat_S\varphi\}$ for all $x$. Also, by the semantics of $\varphi\to\psi$, $\neg\varphi$ and $\bot$ we deduce that $des_S^\alpha(x)$ is an $S$-theory, negation complete, and $\bot$-free.
	\end{proof}
	
	The $S$-description sets satisfy some further properties that we now define in order to characterize them:
	
	\begin{defi}
		Given a $T$-deduction system $\{\ded_S\st S\in\ing T\}$, a set $\Gamma\subseteq\mathsf{Form}_S$, is $\ded_S$-\emph{consistent} if $\Gamma\not\ded_S\bot_S$, and $\ded_S$-\emph{inconsistent} otherwise. If $\Gamma$ is negation complete and $\ded_S$-consistent, we say $\Gamma$ is $\ded_S$-\emph{maximal}.  
		
		\label{deflinda}
		A $T$-deduction system is \emph{Lindenbaum} if for all $S\in\ing T$, every $\ded_S$-consistent set of formulas is included in some $\ded_S$-maximal set.
	\end{defi}
	
	Some of the properties of $\ded_S$-maximal sets are compiled in the following lemma:
	
	\begin{lemC}[\cite{goldblattdeduction2010}]\label{tautologiasenmaximales}\ 
		\begin{enumerate}[\normalfont(1)]
			\item $\Gamma$ is $\ded_S$-maximal iff is $\ded_S$-consistent, and has no proper extensions that are $\ded_S$-consistent.\label{propmaximalnoproperextension}
			\item If $\Gamma$ is $\ded_S$-maximal, then\label{propmaximalentonces}
			\begin{enumerate}[\normalfont(a)]
				\item $\Gamma$ is $\ded_S$-closed, that is, $\Sigma\subseteq\Gamma$ and $\Sigma\ded_S\varphi$ implies $\varphi\in\Gamma$.\label{propcjtocerradoform}
				\item $\ded_S\varphi$ implies $\varphi\in\Gamma$.\label{propaxiomaspertenecenamaximal}
			\end{enumerate}
			\item If a $T$-deduction system is Lindenbaum, then $\ded_S\varphi$ iff for every $\ded_S$-maximal set $\Gamma$, $\varphi\in\Gamma$.
			\label{proplindequivpertenacjtomax}
		\end{enumerate}
	\end{lemC}
	
	The last part of the lemma says that in a Lindenbaum deduction system $D$, we can relate the notion of a formula $\varphi$ being valid syntactically with the fact that such formula belongs to every $\ded_S$-maximal set of formulas.
	
	We now come to a key theorem for the proof of completeness.
	
	\begin{thmC}[\cite{goldblattdeduction2010}]\label{teolinda}
		For any $T$-coalgebra $\alpha$, in a given $T$-deduction system:
		\begin{enumerate}[\normalfont(1)]
			\item Every satisfiable set of $S$-formulas is $\ded_S$-consistent.\label{teolindasoundiffsatisfiableconsistent}
			\item Every $S$-description set is $\ded_S$-maximal.\label{teolindasounddesismaximal}
			\item $Conseq_T^\alpha=\{\sat_S^\alpha\st S\in\ing T\}$ and $Conseq_T=\{\sat_S\st S\in\ing T\}$ are Lindenbaum $T$-deduction systems.\label{teolindaconseq}
		\end{enumerate}
	\end{thmC}
	
	The proof of the Lindenbaum properties for the systems $Conseq_T^\alpha$ and $Conseq_T$ appeal to the description sets of elements in coalgebras, which are used in \cite{moss04harsanyi} to build the canonical coalgebras.
	
	\section{Canonical coalgebras and completeness}\label{seccanonicalspacesandcoalgebras}
	
	Our goal in this section is to prove the completeness of the coalgebraic logic defined for each upper probability polynomial functor $T$. We do this by using the method of constructing a canonical coalgebra made of $\ded_S$-maximal sets. As intermediate steps, we will need to build uncertainty spaces out of collections of $\ded_S$-maximal sets and functions connecting them. We are adapting the methods from \cite{moss04harsanyi} and \cite{goldblattdeduction2010} to our setting to account for the upper and lower probability measures. Along the way we will cite theorems from those works that can be carried without changes, and focus on the new results needed for this particular generalization.
	
	\begin{defi}\label{elementoscanonicos}
		For a Lindenbaum $T$-deduction system $D$, and each functor $S\in\ing(T)$ we define the pair $(X^D_S,\Sigma_S^D)$, where $X_S^D=\{x\subseteq\mathsf{Form}_S\st x\text{ is }\ded_S\text{-maximal}\}$, $\Sigma_S^D=\{\vabs{\varphi}_S\st\varphi::S\}$, and $\vabs{\varphi}_S=\{x\in X_S^D\st\varphi\in x\}$.
	\end{defi}
	
	Since by Lemma \ref{proptheories} $\vabs{\varphi\lor\psi}_S=\vabs{\varphi}_S\cup\vabs{\psi}_S$ and $\vabs{\neg\varphi}_S=X_S^D\setminus\vabs{\varphi}_S$, $\Sigma_S^D$ is an algebra of sets, so we don't need here to take the algebra generated by the sets of the form $\vabs{\varphi}_S$. 
	
	We assume that the deduction system $D$ in this section has the Lindenbaum property. This is crucial to prove:
	
	\begin{lemC}[\cite{goldblattdeduction2010}]\label{relacinterpaxiom} 
		For any $\varphi,\psi:S$, $\vabs{\varphi}_S\subseteq\vabs{\psi}_S$ iff $\ded_S\varphi\to\psi$, and \label{10102021-1} $\vabs{\varphi}_S=\vabs{\psi}_S$ iff $\ded_S\varphi\leftrightarrow\psi$.\label{10102021-2}
	\end{lemC}
	
	\begin{defi}
		\label{canonicalmeasure}
		If $\Delta^*S\in\ing T$, for each $\ded_{\Delta^*S}$-maximal set $x\in X_{\Delta^*S}^D$, and $\varphi::S$,
		\[g_x(\vabs{\varphi}_S)=\sup\{q\in[0,1]\st\upper{\ge q}\varphi\in x\}.\]
	\end{defi}
	Notice here that even if the values of $q$ are restricted to the rationals, $g_x$ can take any real value.
	
	\begin{lem}\label{propmedcan}
		For each $x\in X_{\Delta^*S}^D$
		\begin{enumerate}[\normalfont(1)]
			\item $g_x(\vabs{\varphi}_S)<p$ implies $\neg\upper{\ge p}\varphi\in x$.\label{propmedcan1}
			\item $\upper{\le p}\varphi\in x$ implies $g_x(\vabs{\varphi}_S)\le p$.\label{propmedcan2}
		\end{enumerate}
	\end{lem}
	
	\begin{proof} 
		To prove \ref{propmedcan1}, suppose that  $\neg\upper{\ge p}\varphi\notin x$. Since $x$ is negation complete, this implies that $\upper{\ge p}\varphi\in x$, so by definition of $g_x(\vabs{\varphi}_S)$ we conclude that $ g_x(\vabs{\varphi}_S)\ge p$, and the result follows by contraposition.
		
		For \ref{propmedcan2} we prove the contrapositive: assume that $g_x(\vabs{\varphi}_S)>p$. By the definition of supremum, there exists some $q\in[0,1]$ such that $q>p$ and (i) $\upper{\ge q}\varphi\in x$. If it were the case that $\upper{\le p}\varphi\in x$, then by axiom \ref{axupprob}\ref{axupprob1} and Detachment it follows that $\upper{<q}\varphi\in x$, or equivalently, $\neg\upper{\ge q}\varphi\in x$, which contradicts (i), since $x$ is $\ded_{\Delta^*S}$-maximal.
	\end{proof}
	
	\begin{thm}\label{canonicalmeasurebuenadef}
		$g_x$ is a well defined upper probability measure on $\Sigma_S^D$.
	\end{thm}
	
	\begin{proof}
		First of all we need to check that the definition of $g_x$ is independent from the formula $\varphi::S$ that designates the set $\vabs{\varphi}_S$. Indeed, if $\psi::S$ and $\vabs{\varphi}_S=\vabs{\psi}_S$ then by Lemma \ref{relacinterpaxiom} $\ded_S\varphi\leftrightarrow\psi$, so by Lemma \ref{dedproplema} \ref{prop5bis}, $\ded_{\Delta^*S}\upper{\ge q}\varphi\leftrightarrow\upper{\ge q}\psi$ for any $q\in [0,1]$, and then $\upper{\ge q}\varphi\in x$ if and only if $\upper{\ge q}\psi\in x$.
		
		By Lemma \ref{dedproplema} \ref{prop1} we have that $0\in\{q:\upper{\ge q}\varphi\in x\}$, and so  $g_x(\vabs{\varphi}_S)$ is a supremum of a nonempty subset of $[0,1]$.
		
		To check the condition \ref{30032024-1}, notice that $\emptyset=\vabs{\bot_S}_S$. By Lemma \ref{dedproplema} \ref{prop3}, $\upper{< p}\bot_S\in x$ for every $p>0$. This is equivalent to $\neg\upper{\ge p}\bot_S\in x$ if $p>0$, which is to say that $\upper{\ge p}\bot_S\notin x$ if $p>0$, so we have that $g_x(\emptyset)=0$.
		
		For \ref{30032024-2} consider $\ded_S\top_S$. By the necessitation rule, it follows that $\ded_{\Delta^*S}\plower{\ge1}\top_S$. By Lemma \ref{dedproplema} \ref{prop2bis} and detachment, we obtain $\ded_{\Delta^*S}\upper{\ge1}\top_S$, so $\upper{\ge1}\top_S\in x$. Therefore, $g_x(\vabs{\top_S}_S)=1$.
		
		For \ref{30032024-3}, we analyse first the case in which $n>0$ and $g_x(\vabs{\varphi}_S)>0$. For some $\vabs{\varphi}_S$, consider a finite sequence $\vabs{\varphi_1}_S,\ldots,\vabs{\varphi_m}_S$ that is a $(n,k)$-cover of $(\vabs{\varphi}_S,\vabs{\top}_S)$. This means that
		\begin{equation}\label{10102021-5}
			\vabs{\top}_S\subseteq\bigcup_{\substack{I\subseteq J_m\\ \vabs{I}=n+k}}\bigcap_{i\in I}\vabs{\varphi_i}_S,
		\end{equation} and
		\begin{equation}\label{10102021-6}
			\vabs{\varphi}_S\subseteq\bigcup_{\substack{I\subseteq J_m\\ \vabs{I}=n}}\bigcap_{i\in I}\vabs{\varphi_i}_S.
		\end{equation}
		
		From \ref{10102021-5}, it follows that (i) $\ded_S\bigvee_{I\subseteq J_m,\vabs{I}=n+k}\bigwedge_{i\in I}\varphi_i$. On the other hand, from \ref{10102021-6} and Lemma \ref{relacinterpaxiom}, we have that (ii) $\ded_S\varphi\to\bigvee_{I\subseteq J_m,\vabs{I}=n}\bigwedge_{i\in I}\varphi_i$, so by (i), (ii) and Lemma \ref{tautologiasenmaximales} \ref{proplindequivpertenacjtomax}, those formulas belong to every maximal set of sort $S$.
		
		Suppose towards contradiction that
		\begin{equation}\label{10102021-3}
			g_x(\vabs{\varphi_1}_S)+\ldots+g_x(\vabs{\varphi_m}_S)<ng_x(\vabs{\varphi}_S)+k.
		\end{equation}
		Let  $0<d=\displaystyle ng_x(\vabs{\varphi}_S)+k-\sum_{i=1}^mg_x(\vabs{\varphi_i}_S).$
		
		Notice that we cannot have $g_x(\vabs{\varphi_i}_S)=1$ for every $1\le i\le m$, because in that case we would have $m<ng_x(\vabs{\varphi}_S)+k\le n+k\le m$. Now let $\epsilon=d/2m$, and if $g_x(\vabs{\varphi_i}_S)<1$, let $\epsilon_i=\epsilon \land (1-g_x(\vabs{\varphi_i}_S))/2$. Let $p_i$ be 1 if $g_x(\vabs{\varphi_i}_S)=1$ and $g_x(\vabs{\varphi_i}_S)+\epsilon_i$ otherwise. Define $q=\frac{\sum_{i=1}^mp_i-k}{n}$, and let $p=0\lor(q\land1)$. Thus we have that (iii) $0\le p<g_x(\vabs{\varphi}_S)\le 1$.  
		
		If $g_x(\vabs{\varphi_i}_S)=1$, then $g_x(\vabs{\varphi_i}_S)\le p_i$, and since Lemma \ref{dedproplema} \ref{prop1} ensures that $\ded_{\Delta^*S}\plower{\ge0}\neg\varphi_i$, we deduce that (iv) $\ded_{\Delta^*S}\upper{\le1}\varphi_i$. On the other hand, if $g_x(\vabs{\varphi_i}_S)<1$, we have that $g_x(\vabs{\varphi_i}_S)<p_i$, and by Lemma \ref{propmedcan} \ref{propmedcan1}, we have that $\neg\upper{\ge p_i}\varphi_i\in x$, which is equivalent to $\upper{<p_i}\varphi_i\in x$, and using axiom \ref{axupprob}\ref{axupprob2} it follows that (v) $\upper{\le p_i}\varphi_i\in x$. By (iv) and (v), we have that (vi) $\upper{\le p_i}\varphi_i\in x$ for all $i$.
		
		By the assumption and cover rules, we can deduce that (vii) $(\upper{\le p_1}\varphi_1\land\cdots\land\upper{\le p_m}\varphi_m)\to\upper{\le p}\varphi\in x$.
		
		Now, by (vi) it follows that (viii) $\upper{\le p_1}\varphi_1\land\cdots\land\upper{\le p_m}\varphi_m\in x$. Therefore, (vii) and (viii) imply that $\upper{\le p}\varphi\in x$. By Lemma \ref{propmedcan} \ref{propmedcan2}, we deduce that $g_x(\vabs{\varphi}_S)\le p$, which contradicts (iii), so \ref{30032024-3} is valid in this case.
		
		If $n=0$ or $g_x(\vabs{\varphi}_S)=0$, we need to show that $k\le \sum_{i=1}^mg_x(\vabs{\varphi_i}_S)$. Suppose that $\sum_{i=1}^mg_x(\vabs{\varphi_i}_S)<k$. Then we can define as above numbers $p_i\in[0,1]$ such that (ix) $\sum_{i=1}^mg_x(\vabs{\varphi_i}_S)\le\sum_{i=1}^mp_i<k$, and $g_x(\vabs{\varphi_i}_S)\le p_i$ for all $i$. Also in this case we cannot have all $g_x(\vabs{\varphi_i}_S)=1$. 
		Notice also that (i) is true in this case too (equation \ref{10102021-5} holds with $n=0$), so we use (i), (ix) and cover rule \ref{nkcoverrule2}, to conclude that $\ded_{\Delta^*S}\neg(\upper{\le p_1}\varphi_1\land\cdots\land\upper{\le p_m}\varphi_m)$, then $\neg(\upper{\le p_1}\varphi_1\land\cdots\land\upper{\le p_m}\varphi_m)\in x$. This means that $\neg\upper{\le p_1}\varphi_1\lor\cdots\lor\neg\upper{\le p_m}\varphi_m\in x$, hence (x) $\neg\upper{\le p_j}\varphi_j\in x$ for some $j$. This $j$ is such that $p_j<1$ because if $p_j=1$, then both $\neg\upper{\le 1}\varphi_j$ and $\upper{\le 1}\varphi_j$ are in $x$, contradicting the fact that $x$ is consistent. Therefore, for this $j$, (x), axiom 
		\ref{axupprob}\ref{axupprob2} and Detachment it follows that $\upper{\ge p_j}\varphi_j\in x$, so by definition of supremum, $g_x(\vabs{\varphi_j}_S)\ge p_j$, which contradicts (ix).
	\end{proof}

	Now that we know that $g_x$ is an upper probability measure, we can prove further results about it.
	
	\begin{lem}\label{relacionmedcanocicaconformulas}
		For all $x\in X_{\Delta^*S}^D$, $\vabs{\varphi}_S\in\Sigma_S^D$ and $p,q\in[0,1]:$
		\begin{enumerate}[\normalfont(1)]
			\item $g_x(\vabs{\varphi}_S)\ge p$ iff $\upper{\ge p}\varphi\in x$,\label{23102021-1}
			\item $\underline{g_x}(\vabs{\varphi}_S)=1-g_x(\vabs{\neg\varphi}_S)\ge q$ iff $\plower{\ge q}\varphi\in x$,\label{23102021-2}
			\item $(g_x(\vabs{\varphi}_S)\ge p$ and $\underline{g_x}(\vabs{\varphi}_S)\ge q)$ iff $[p,q]\varphi\in x$.\label{23102021-3}
		\end{enumerate}
	\end{lem}
	
	\begin{proof}
		\ref{23102021-1} The right-to-left implication follows directly by definition of $g_x$.
		
		Conversely, if $g_x(\vabs{\varphi}_S)\ge p$, then (i) for all $p'<p$, $\upper{\ge p'}\varphi\in x$. Indeed, since (ii) $g_x(\vabs{\varphi}_S)>p'$, if it were the case that $\upper{\ge p'}\varphi\notin x$, then by Lemma \ref{dedproplema} \ref{prop7}, for any $q>p'$, $\upper{\ge q}\varphi\notin x$. Thus, if $\upper{\ge q}\varphi\in x$ then $q\le p'$, and therefore $p'$ is an upper bound for $g_x(\vabs{\varphi}_S)$, but this contradicts (ii).
		
		Now that we have proved (i), by the Archimedean Rule and the fact that $x$ is $\ded_S$-closed, we obtain $\upper{\ge p}\varphi\in x$.
		
		\ref{23102021-2} From right to left, consider the equivalence $\plower{\ge q}\varphi\in x$ iff $\upper{\le 1-q}\neg\varphi\in x$, which holds by Observation \ref{changeofnotation}. By Lemma \ref{propmedcan} \ref{propmedcan2}, it follows that $g_x(\vabs{\neg\varphi}_S)\le 1-q$, or equivalently $\underline{g_x}(\vabs{\varphi}_S)=1-g_x(\vabs{\neg\varphi}_S)\ge q$.
		
		Conversely, suppose that (iii) $g_x(\vabs{\neg\varphi}_S)\le 1-q$ and $q'<q$. Assume towards contradiction that $\plower{\ge q'}\varphi\notin x$, so $\neg\plower{\ge q'}\varphi\in x$ which is equivalent to $\neg\upper{\le 1-q'}\neg\varphi\in x$. By axiom \ref{axupprob}\ref{axupprob2}, $\upper{<1-q'}\neg\varphi\to\upper{\le 1-q'}\neg\varphi\in x$, so by contraposition, it follows that $\neg\upper{<1-q'}\neg\varphi\in x$, this is $\upper{\ge 1-q'}\neg\varphi\in x$. By the definition of $g_x$, it follows that $1-q'\le g_x(\vabs{\neg\varphi}_S)\underset{\text{(iii)}}{\le}1-q$, so $q\le q'$, which contradicts the hypothesis. We then conclude that $\plower{\ge q'}\varphi\in x$ for all $q'<q$, and by the Archimedean Rule, $\plower{\ge q}\varphi\in x$.
		
		\ref{23102021-3} It is a direct consequence of parts \ref{23102021-1}, \ref{23102021-2}, and axiom \ref{axupprob}\ref{axupprob6}.
	\end{proof}
	
	\
	
	Since $D$ is a fixed $T$-deduction system, we will from now on denote the spaces $X^D_S$ simply as $X_S$.
	
	We need to find the relations between the $\ded_S$-maximal sets of different ingredients, and for this we define for each $\kappa\ne(p,q)$ with $S\overset{\kappa}{\rightsquigarrow}S'$ and $\Gamma:S$, the set $[\kappa]^-\Gamma=\{\varphi:S'\st[\kappa]\varphi\in\Gamma\}$. Then, if $\Gamma$ is a $\ded_S$-maximal set, and $\kappa$ is of the form $\pr_j$, $\ev_e$ or $\sig$, the set $[\kappa]^-\Gamma$ is $\ded_S$-maximal. In the case of $\kappa=\inc_j$, $\neg[\inc_j]\bot_{S_j}\in\Gamma$ implies that $[\inc_j]^-\Gamma$ is $\ded_{S_j}$-maximal (see \cite{goldblattdeduction2010} for the details).
	
	\begin{lem}\label{measurablefunctions} There exist functions:
		\begin{enumerate}[\normalfont(1)]
			\item $\rho_{S_1\times S_2}:X_{S_1\times S_2}\to X_{S_1}\times X_{S_2}$, defined by $\rho_{S_1\times S_2}(x)=([\pr_1]^-x,[\pr_2]^-x)$ for every $x\in X_{S_1\times S_2}$, that satisfies $\rho^{-1}_{S_1\times S_2}(\vabs{\varphi_1}_{S_1}\times\vabs{\varphi_2}_{S_2})=\vabs{[\pr_1]\varphi_1\land[\pr_2]\varphi_2}_{S_1\times S_2}$.
			\item $\rho_{S_1+S_2}:X_{S_1+S_2}\to X_{S_1}+X_{S_2}$, defined by $\rho_{S_1+S_2}(x)=in_j([\inc_j]^-x)$ for every $x\in X_{S_1+S_2}$, where $j\in\{1,2\}$ is the only index such that $\neg[\inc_j]\bot_{S_j}\in x$, that satisfies $\rho^{-1}_{S_1+S_2}(in_j(\vabs{\varphi}_{S_j}))=\vabs{\neg[\inc_j]\bot_{S_j}\land[\inc_j]\varphi}_{S_1+S_2}$.
			\item $\rho_{Id}:X_{Id}\to X_{T}$, defined by $\rho_{Id}(x)=[\sig]^-x$ for every $x\in X_{Id}$, that  satisfies $\rho^{-1}_{Id}(\vabs{\varphi}_{T})=\vabs{[\sig]\varphi}_{Id}$.
			\item $\rho_{\Delta^*S}:X_{\Delta^*S}\to \Delta^*(X_{S})$, defined by $\rho_{\Delta^*S}(x)=g_x$ for every $x\in X_{\Delta^*S}$, where $g_x$ is the upper probability measure from Theorem \ref{canonicalmeasurebuenadef}, that  satisfies $\rho^{-1}_{\Delta^*S}(\beta^{p,q}\vabs{\varphi}_{S})=\vabs{[(p,q)]\varphi}_{\Delta^*S}$ for every $\varphi::S$.
		\end{enumerate}
		It follows that these functions are all measurable.
	\end{lem}
	
	\begin{proof} The measurability of the functions follows in each case from the satisfied condition, since the sets of the form $\vabs{\varphi}_S$, with $\varphi::S$ are all the elements of the algebra. These conditions are proved as in \cite{goldblattdeduction2010} and \cite{moss04harsanyi}, except for the case of $\rho_{\Delta^*}$: for this we use Lemma \ref{relacionmedcanocicaconformulas}, and the result follows from the equivalences $\rho_{\Delta^*S}(x)\in\beta^{p,q}(\vabs{\varphi}_S)$ iff ($g_x(\vabs{\varphi}_S)\ge p$ and $\underline{g_x}(\vabs{\varphi}_S)\ge q$) iff ($\upper{\ge p}\varphi\in x$ and $\plower{\ge q}\varphi\in x$) iff $[(p,q)]\varphi\in x$ iff $x\in\vabs{[(p,q)]\varphi}_{\Delta^*S}$.
	\end{proof}
	
	We have now built the space $X_{Id}$ of all the $\ded_{Id}$-maximal sets, to which one can apply the functors $S\in\ing T$. In order to establish the connection between the spaces $S(X_{Id})$ and $X_S$, we introduce the following subsets of $S(X_{Id})$:
	
	\begin{defi}
			\begin{itemize}
				\item $\interp{\bot_S}_S=\emptyset$,
				\item $\interp{A}_M=A$,
				\item $\interp{\varphi_1\to\varphi_2}_S=(S(X_{Id})\setminus\interp{\varphi_1}_S)\cup\interp{\varphi_2}_S$,
				\item $\interp{[\pr_j]\varphi}_{S_1\times S_2}=\pi_j^{-1}\interp{\varphi}_{S_j}$,
				\item $\interp{[\inc_1]\varphi}_{S_1+S_2}=in_1(\interp{\varphi}_{S_1})\cup in_2(S_2(X_{Id}))$,
				\item $\interp{[\inc_2]\varphi}_{S_1+S_2}=in_1(S_1(X_{Id}))\cup in_2(\interp{\varphi}_{S_2})$,
				\item $\interp{[\sig]\varphi}_{Id}=\vabs{[\sig]\varphi}_{Id}$,
				\item $\interp{[(p,q)]\varphi}_{\Delta^*S}=\beta^{p,q}\interp{\varphi}_S$, if $\varphi::S$.
			\end{itemize}
	\end{defi}
	
	It follows from the definition that for every $\varphi::S$, $\interp{\varphi}\in \Sigma_{S(X_{Id})}$. With these sets in hand, we can define the maps $\ere_S$ that realize the connection:
	
	\begin{lem}\label{measurablemapR}
		There are measurable maps $\ere_S:X_S\to S(X_{Id})$ such that for every $S\in\ing T$ and $\varphi:S$  we have that $\ere_S^{-1}\interp{\varphi}_S=\vabs{\varphi}_S$, which is equivalent to $\rm{(1)}$ $\ere_S(x)\in\interp{\varphi}_S$ iff $\varphi\in x$. 
	\end{lem}

	\begin{proof}
		For each ingredient $S$, the set of formulas $\varphi$ that satisfy the condition $(1)$ include $\bot_S$ and is closed under implication. The maps $\ere_S$ are inductively defined as follows:
		
		\begin{itemize}
			\item $\ere_{Id}:X_{Id}\to X_{Id}$ is the identity map. The condition $\ere_{Id}^{-1}\interp{\varphi}_{Id}=\vabs{\varphi}_{Id}$ follows from the fact that $\interp{\varphi}_{Id}=\vabs{\varphi}_{Id}$, which is proved by induction using Lemma \ref{proptheories}.
			\item For each $x\in X_M$, there exists a unique $c\in M$ such that $\{c\}\in x$. This is a consequence of the Constant Rule, axiom \ref{axconstant}\ref{axconstant2}, and the fact that $x$ is $\ded_M$-consistent. Thus, we can define $\ere_{M}(x)$ to be that element $c$.
			\item Assume that $\ere_{S_1}$ and $\ere_{S_2}$ are defined. We put $\ere_{S_1\times S_2}=(\ere_{S_1}\times\ere_{S_2})\circ\rho_{S_1\times S_2}$. Hence, $\ere_{S_1\times S_2}(x)=(\ere_{S_1}([\pr_1]^-x),\ere_{S_2}([\pr_2]^-x))$.
			\item Similarly, let $\ere_{S_1+S_2}=(\ere_{S_1}+\ere_{S_2})\circ\rho_{S_1+S_2}$, so $\ere_{S_1+S_2}(x)=in_j(\ere_{S_j}([\inc_j]^-x))$, where $j\in \{1,2\}$ is the only index such that $\neg[\inc_j]\bot_{S_j}\in x$.
			\item For the ingredient $\Delta^*S$, let $\ere_{\Delta^*S}$ be the composition $X_{\Delta^*S}\overset{\rho_{\Delta^*S}}{\longrightarrow}\Delta^*(X_S)\overset{\Delta^*\ere_{S}}{\longrightarrow}\Delta^*S(X_{Id})$. Hence, $\ere_{\Delta^*S}(x)(A)=g_x(\ere_S^{-1}(A))$ for all measurable subsets $A$ of $S(X_{Id})$.
		\end{itemize}
		
		We will only prove that $\ere_{\Delta^*S}$ is measurable and satisfies (1) (the measurability in other cases can be proved adapting Lemma 5.7 of \cite{goldblattdeduction2010}).
		
		$\ere_{\Delta^*S}$ is measurable because it is the composition of measurable functions. For the condition (1), $\ere_{\Delta^*S}^{-1}\interp{[(p,q)]\varphi}_{\Delta^*S}=\rho^{-1}_{\Delta^*S}((\Delta\ere_S)^{-1}(\beta^{p,q}\interp{\varphi}_S))$. By Lemma \ref{losbetasmedibles} this is equal to $\rho^{-1}_{\Delta^*S}(\beta^{p,q}(\ere_S^{-1}\interp{\varphi}_S))$, and by inductive hypothesis, this equals $\rho^{-1}_{\Delta^*S}(\beta^{p,q}(\vabs{\varphi}_S))$, which in turn is equal to  $\vabs{[(p,q)]\varphi}_{\Delta^*S}$ by definition.
	\end{proof}
	
	\begin{defi}\label{defcanonicalcoalgebra} The \emph{canonical $T$-coalgebra} for a Lindenbaum $T$-deduction system $D$ is $(X_{Id},\alpha^D)$, where $\alpha^D=\ere_T\circ\rho_{Id}:X_{Id}\to T(X_{Id})$.
	\end{defi}
	
	As it turns out, the set $\interp{\varphi}_S$ will be the semantic interpretation of the formula $\varphi$ in the canonical coalgebra, as the following Lemma shows:
	
	\begin{lemC}[\cite{goldblattdeduction2010}]
		\label{truthlemma}
		For each $\varphi:S$, we have that $\class{\varphi}_S^{\alpha^D}=\interp{\varphi}_S$. Also, for every $\ded_S$-maximal set $x$, the condition $\alpha^D,\ere_S(x)\sat_S\varphi$ is equivalent to $\varphi\in x$.
	\end{lemC}
	
	So far, we have considered the description sets by themselves, but it should be clear that for each $T$-coalgebra $(Y,\alpha)$ and each ingredient $S$ of $T$, $des^\alpha_S$ is a map from $SY$ to $X_S$. These maps have the following properties:
	
	\begin{thm}\label{propertiesdescription}
		\begin{enumerate}[\normalfont(1)]
			\item For every ingredient $S$ and every coalgebra $(Y,\delta)$, $des^\delta_S:SY\to X_S$ is a measurable map.\label{desaremeasurable}
			\item $T$-coalgebra morphisms preserve the description maps: if $f:(X,\alpha)\to(Y,\delta)$ then for all ingredients $S$, $des^\delta_Sf=des^\alpha_S$.\label{desmorphismpreservesemantics}
			\item For the canonical $T$-coalgebra $(X_{Id},\alpha^D)$, $des_S^{\alpha^D}:SX_{Id}\to X_S$ is an isomorphism with inverse $\ere_S$. In particular, $X_S=\{des_S^{\alpha^D}(\ere_S(x))\st x\in X_S\}$. \label{desareisomorphism}
			\item For all $T$-coalgebras $(Y,\delta)$, $des_{Id}^\delta:Y\to X_{Id}$ is a $T$-coalgebra morphism. Furthermore, it is the only morphism, so $(X_{Id},\alpha^D)$ is the final $T$-coalgebra.\label{desfinalcoalgebra}
		\end{enumerate}
	\end{thm}
	\begin{proof}
		
		\ref{desaremeasurable} To prove that $des^{\delta}_S$ is measurable, it is enough to show that $(des^{\delta}_S)^{-1}\vabs{\varphi}_S$ is a measurable set for all $\varphi::S$, but $(des^{\delta}_S)^{-1}\vabs{\varphi}_S=\{y\in SY\st\varphi\in des^{\delta}_S(y)\}=\{y\st\delta,y\sat\varphi\}=\class{\varphi}^{\delta}_S$, and we know that the sets $\class{\varphi}^{\delta}_S$ are measurable.
		
		\ref{desmorphismpreservesemantics} By Lemma \ref{preservation}, $\varphi\in des^\delta_S Sf(x)$ iff $Sf(x)\in\class{\varphi}^\delta_S$ iff $x\in (Sf)^{-1}\class{\varphi}^\delta_S$ iff $x\in \class{\varphi}^\alpha_S$ iff $\varphi\in des^\alpha_S(x)$.
		
		\ref{desareisomorphism} By Lemma \ref{truthlemma} we have that $des_S^{\alpha^D}\circ\ere_S(x) =\set{\varphi:S\st \alpha^D,\ere_S(x)\sat\varphi}=x$, so it suffices to show that $\ere_S\circ des_S^{\alpha^D}=Id_{S(X_{Id})}$. The proof is by induction on the ingredients of $T$. We only show the case for $\Delta^*S\in\ing T$ (the other cases are as in \cite{goldblattdeduction2010}).
		
		Let $g\in\Delta^*S(X_{Id})$ and $x=des^{\alpha^D}_{\Delta^*S}(g)$. By inductive hypothesis we have that $\ere_S:X_S\to S(X_{Id})$ is an isomorphism, and as a consequence, all the measurable sets of $S(X_{Id})$ are of the form $\vabs{\varphi}_S$ for some $\varphi::S$. From Lemma \ref{measurablemapR}, $\ere_{\Delta^*S}(x)(U)=g_x(\ere_S^{-1}(U))$, for all measurable subsets $U$ of $S(X_{Id})$, so this implies that $\ere_{\Delta^*S}(x)$ verifies
		\[\ere_{\Delta^*S}(x)(\ere_S\vabs{\varphi}_S)=g_x((\ere_S)^{-1}\ere_S\vabs{\varphi}_S)=g_x(\vabs{\varphi}_S).\]
		By Lemma \ref{truthlemma}, $\class{\varphi}_S^{\alpha^D}=\interp{\varphi}_S$, and by Lemma \ref{measurablemapR}, $\ere_S^{-1}(\interp{\varphi}_S)=\vabs{\varphi}_S$, so $\vabs{\varphi}_S=\ere_S^{-1}(\class{\varphi}_S^{\alpha^D})$, which is to say that $\ere_S\vabs{\varphi}_S=\class{\varphi}_S^{\alpha^D}$. We have then that $g(\ere_S\vabs{\varphi}_S)=g(\class{\varphi}_S^{\alpha^D})\ge q$ iff $\upper{\ge q}\varphi\in des^{\alpha^D}_{\Delta^*S}(g)=x$. Hence, $g(\ere_S\vabs{\varphi}_S)=\sup\{q\in[0,1]\st\upper{\ge q}\varphi\in x\}=g_x(\vabs{\varphi}_S)$, so $\ere_{\Delta^*S}(x)=g$.
		
		\ref{desfinalcoalgebra} Consider the following diagram:
		\[
		\xymatrix{
			Y\ar[r]^{\delta}\ar[d]_{des^\delta_{Id}}	&TY\ar[d]^{des^\delta_T}\ar[dr]^{Tdes^\delta_{Id}}	\\
			X_{Id}\ar[r]_{\rho_{Id}}& X_{T}\ar[r]_{\ere_{T}}&TX_{Id}
		}
		\]
		The square on the left commutes: for any $y\in Y$, 
		$\rho_{Id}des^\delta_{Id}(y)$ $=$ $[\sig]^{-}\set{\varphi:Id\st \delta,y\sat\varphi}$ $=$ $\set{\psi:T\st[\sig]\psi\in\set{\varphi:Id\st\delta, y\sat\varphi}}=\set{\psi:T\st\delta,y\sat[\sig]\psi}$ $=$ $\set{\psi:T\st\delta,\delta(y)\sat\psi}$ $=$ \linebreak[4]$des^\delta_T(\delta(y))$.

		To prove that the triangle of the right also commutes, we proceed by induction on the definition of $T$, and prove that for every ingredient $S$, $\ere_Sdes^\delta_S=Sdes^\delta_{Id}$. We indicate here the proof for the ingredients of the form $\Delta^*S$. In the following diagram, the triangle on the right commutes by application of $\Delta^*$ to the inductive hypothesis:
		\[
		\xymatrix{
			\Delta^*SY\ar[d]_{des^\delta_{\Delta^*S}}\ar[dr]_{\Delta^*des^\delta_{S}}\ar[drr]^{\Delta^*Sdes^\delta_{Id}}\\
			X_{\Delta^*S}\ar[r]_{\rho_{\Delta^*S}}&\Delta^*X_{S}\ar[r]_{\Delta^*\ere_{S}}&\Delta^*SX_{Id}
		}
		\]
		For the triangle on the left, we want to prove that  for every $g\in\Delta^*SY$ and $\varphi::S$, \linebreak[4]$\rho_{\Delta^*S}des^\delta_{\Delta^*S}(g)(\vabs{\varphi})=$ $\Delta^*des^\delta_{s}(g)(\vabs{\varphi})$. Let $x=des^\delta_{\Delta^*S}(g)$. Then $\rho_{\Delta^*S}des^\delta_{\Delta^*S}(g)(\vabs{\varphi})=$ \linebreak[4] $\rho_{\Delta^*S}(x)(\vabs{\varphi})=g_x(\vabs{\varphi})=sup\set{q\st\upper{\ge q}\varphi\in x}=sup\set{q\st \upper{\ge q}\varphi\in des^\delta_{\Delta^*S}(g)}=$\linebreak[4] $sup\set{q\st g(\class{\varphi}^\delta_S){\ge q}}=g(\class{\varphi}^\delta_S)$. By the first part of this theorem, this equals  
		$g(des^\delta_S)^{-1}(\vabs{\varphi})$ $=\Delta^*des^\delta_{s}(g)(\vabs{\varphi})$.
		
		Finally, to prove that $des^\delta_{Id}$ is the only $T$-coalgebra morphism from $Y$ to $X_{Id}$, notice first that $des^{\alpha^D}_{Id}$ is the identity function $1_{X_{Id}}$: for any $x\in X_{Id}$, $des^{\alpha^D}_{Id}(x)$ is by Lemma \ref{truthlemma} the set $\set{\varphi:Id\st x\in|\varphi|_{Id}}=\set{\varphi:Id\st \varphi\in x}=x$. Then, if $f:(Y,\delta)\to(X_{Id},\alpha^D)$ is a morphism, we have by item \ref{desmorphismpreservesemantics} that $des^{\alpha^D}_{Id}f=des^\delta_{Id}$, but since $des^{\alpha^D}_{Id}=1_{X_{Id}}$, this means that $f=des^\delta_{Id}$.
	\end{proof}
	
	\begin{defi}\label{defcomplete}
		A $T$-deduction system $\{\ded_S\st S\in\ing T\}$ is \emph{complete} if, for all $S$ and all $\Gamma\cup\{\varphi\}\subseteq\mathsf{Form}_S$, $\Gamma\sat_S\varphi$ implies $\Gamma\ded_S\varphi$.
	\end{defi}
	
	\begin{lem}
		A $T$-deduction system is complete iff every $\ded_S$-consistent set of formulas is satisfiable in some $T$-coalgebra.
	\end{lem}
	
	\begin{proof}
		Suppose first that a $T$-deduction system is complete. Let $\Gamma$ be a $\ded_S$-consistent set. Then $\Gamma\not\ded_S\bot_S$, and by hypothesis, this implies that $\Gamma\not\sat_S\bot_S$. Then, there exists a coalgebra $\alpha$ such that $\Gamma\not\sat_S^\alpha\bot_S$. This means that $\Gamma$ is satisfiable.
		
		Conversely, suppose that each $\ded_S$-consistent set of formulas is satisfiable in some $T$-coalgebra. Consider a set $\Gamma$ such that (i) $\Gamma\sat_S\varphi$ and (ii) $\Gamma\not\ded_S\varphi$. By (ii) and Lemma \ref{propsistdeduct} \ref{modalsdprop2}, we have that $\Gamma\cup\{\neg\varphi\}$ is consistent, so by hypothesis, $\Gamma\cup\{\neg\varphi\}$ is satisfiable in some coalgebra $\alpha$, which means that $\sat_S^\alpha\Gamma\cup\{\neg\varphi\}$, so (iii) $\sat_S^\alpha\Gamma$, and (iv) $\sat_S^\alpha\neg\varphi$.
		Also, condition (i) implies that (v) for all coalgebras $\delta$, if $\sat_S^{\delta}\Gamma$ then $\sat_S^{\delta}\varphi$. Combining conditions (iii) and (v), it follows that $\sat_S^\alpha\varphi$, which contradicts (iv).
	\end{proof}
	
	From this Lemma and Theorem \ref{teolinda} \ref{teolindasoundiffsatisfiableconsistent} it follows that in a complete $T$-deduction system, the notions of satisfiability and $\ded_S$-consistency are equivalent.
	
	\begin{thm}[Completeness]\label{completeness}
		For any \emph{Lindenbaum} $T$-deduction system $D$, $\Gamma\sat^{\alpha^D}_S\varphi$ iff $\Gamma\ded_S^D\varphi$.
	\end{thm}
	
	\begin{proof}
		This theorem is proved in the standard way. First, assuming that $\Gamma\not\ded_S^D\varphi$, one can deduce, by the Lindenbaum property, the existence of $x\in X_S$ such that $\Gamma\cup\set{\neg\varphi}\subseteq x$, and by Lemma \ref{truthlemma}, $\alpha^D,\ere_S(x)\sat\Gamma$ while $\alpha^D,\ere_S(x)\not\sat\varphi$, proving that $\Gamma\not\sat_{\alpha^S}^D\varphi$.
		
		In the other direction, if we assume that $\Gamma\not\sat_S^{\alpha^D}\varphi$, then there exists $y\in SX_{Id}$ such that $\alpha^D,y\sat\Gamma$ and $\alpha^D,y\not\sat\varphi$. By Theorem \ref{propertiesdescription} \ref{desareisomorphism}, $y=\ere_S(x)$ for some $x\in X_S$, and by Lemma \ref{truthlemma} again, $\Gamma\subseteq x$ while $\varphi\notin x$. Since $x$ is $\ded_S$-maximal, if it were the case that $\Gamma\ded_S^D\varphi$, then by Lemma \ref{tautologiasenmaximales} \ref{propmaximalentonces}\ref{propcjtocerradoform}, we would have $\varphi\in x$, a contradiction. Thus $\Gamma\not\ded^{D}_S\varphi$.
	\end{proof}
	
	\begin{cor}\label{coroComplete}
		If $D$ is any Lindenbaum $T$-deduction system, the following are equivalent: $\rm{(1)}$ $\Gamma\sat_S\varphi$, $\rm{(2)}$ 
		$\Gamma\sat_S^{\alpha^D}\varphi$,
		and	$\rm{(3)}$ $\Gamma\ded_S^D\varphi$.
	\end{cor}
	
	\begin{proof}
		The statements (2) and (3) are equivalent by Theorem \ref{completeness}. Since $D$ is sound, we have that (3) implies (1), and by definition of $\sat_S$, it follows that (1) implies (2).
	\end{proof}
	
	\section{Other functors for representing uncertainty}
	
	In this section we extend the class of polynomial functors for which the main results are valid, adding the closure under other three functors that can be used to represent uncertainty. For each one of these representations of uncertainty we give the basic properties, check that they are functorial in the category of uncertainty spaces, present modal operators, their semantics, give axioms and inference rules for the corresponding deduction systems, prove their soundness and prove that the canonical measures built with the corresponding formulas are indeed in the corresponding space. Since much of this work is similar to what we have done for upper and lower measures, we include only the proof of the parts that are significatively different.
	
	\subsection{Probability measures}\label{secprobmeasures}
	
	We already mentioned in Definition \ref{defmedprobfinita}, the notion of (finitely additive) probability measures, the most generally used representation for uncertainty in mathematical models.  We will consider the functor $\Delta$ that, given an uncertainty space $X$, returns a measurable space $\Delta X$ consisting of all the probability measures on $X$. This functor has of course been considered before (see \cite{moss04harsanyi} and  \cite{goldblattdeduction2010} for the case of $\sigma$-additive probability measures), but we want here to capitalize the work we have done for upper probability measures, and get a new set of axioms for finitely additive, in a way that is compatible with the setting of the previous section.
	
	In detail, for an uncertainty space $(X,\Sigma_X)$ of $\meas$, the image of $X$ under the functor $\Delta$ is the uncertainty space $(\Delta X,\Sigma_{\Delta X})$, where $\Delta X$ is the space of all (finitely additive) probability measures, and $\Sigma_{\Delta X}$ is the algebra generated by the sets of the form \[\beta_{pr}^p(U)=\set{\mu\in\Delta X\st \mu(U)\ge p},\]
	
	\begin{lem}\label{deltaesfuntorial}
		If $f:X\to Y$ is a  measurable function, defining  $\Delta f$  by $\Delta f(\mu)(U)=\mu(f^{-1}(U))$ for every $\mu\in\Delta X$ and $U\in \Sigma_{Y}$, we have that 
		$(\Delta f)(\mu)$ is a probability measure and $\Delta f: \Delta X\to \Delta Y$ is a measurable function. As a consequence, $\Delta$ is indeed an endofunctor in the category of uncertainty spaces.
	\end{lem}
	
	\begin{proof}
		To see that $\Delta f(\mu)$ is a probability measure over $X'$ it suffices to show that $(\Delta f)(\mu)(U\cup V)=(\Delta f)(\mu)(U)+(\Delta f)(\mu)(V)$ for all disjoint $U,V\in\Sigma$.
		
		Since $f$ is measurable and $\mu$ is a probability measure, we have that $(\Delta f)(\mu)(U\cup V)= \mu(f^{-1}(U\cup V))=\mu(f^{-1}(U)\cup f^{-1}(V))=\mu(f^{-1}(U))+\mu(f^{-1}(V))=(\Delta f)(\mu)(U)+(\Delta f)(\mu)(V)$. Here we have used the fact that the inverse image of a function preserves the disjoint unions of sets.
		
		In an analogous way to Lemma \ref{losbetasmedibles}, one proves that for every  $U\in\Sigma_{X'}$, $(\Delta f)^{-1}(\beta^p_{pr}(U))$ $=\beta_{pr}^p(f^{-1}(U))$, so $\Delta f$ is measurable.
	\end{proof}
	
	We add the labeled edges $\Delta X\overset{\ge p}{\rightsquigarrow}X$ (for each $p\in[0,1]$) to the multigraph of ingredients, and write the associated modal operators as $\proba{\ge p}$.
	
	\begin{obs}\label{whenupprobareprob}
		If $g$ is an upper probability defined on an algebra of sets $\Sigma$, we know by Observation \ref{subaditividaddenkcover} that it is subadditive, that is:
		\begin{equation}\label{26122021}
			\text{for all disjoint }U,V\in\Sigma\text{, }g(U\cup V)\le g(U)+g(V).
		\end{equation}
		Similarly, from the fact that $\underline{g}(U)=1-g(U^c)$, it can be shown that $\underline{g}$ is superadditive. Now, if an upper probability measure $g$ is such that for all $U\in\Sigma$, $g(U)=\underline{g}(U)$, then it is a (finitely additive) probability measure. Indeed, since $g$ verifies \ref{30032024-1} and \ref{30032024-2}, it suffices to show the finite additivity condition. Let $U,V\in\Sigma$ disjoint, then $g(U)+g(V)=\underline{g}(U)+\underline{g}(V)\le\underline{g}(U\cup V)=g(U\cup V)$, which together with \ref{26122021} gives us the additivity.
	\end{obs}
	
	Based on the observation above, one could argue that for the ingredients of the form $\Delta S$, we could repeat the axioms given for $\Delta^*S$ and add the formula
	\[
	\upper{\ge p}\varphi\leftrightarrow\plower{\ge p}\varphi
	\]
	as an axiom to obtain a characterization of finitely additive probabilities. While this is true, we can do better, by writing  all the modal operators in terms of $\proba{\ge p}$. That is:
	\begin{itemize}
		\item Write $\proba{\ge p}\varphi$ instead of $\upper{\ge p}\varphi$ and $\plower{\ge p}\varphi$.
		\item Write $\proba{<p}\varphi$, $\proba{\le p}\varphi$, and $\proba{>p}\varphi$ instead of $\neg\proba{\ge p}\varphi$, $\proba{\ge 1-p}\neg\varphi$, and $\neg\proba{\ge 1-p}\neg\varphi$ respectively.
	\end{itemize}
	Then the analog of axiom \ref{axupprob}\ref{axupprob6} turns out not to be necessary in this setting. We add  the following axioms and deduction rules to the Definition \ref{defdedsystem} of $T$-deduction systems:
	
	\begin{defi} \label{defdeductivesystemprob}
		\begin{enumerate}[\normalfont(1)]\setcounter{enumi}{7}
			\item For an ingredient of the form $\Delta S$,\label{axprob}
			\begin{enumerate}[\normalfont(a)]
				\item $\proba{\le p}\varphi\to\proba{<q}\varphi$ for $p<q$,\label{axprob1}
				\item $\proba{<p}\varphi\to\proba{\le p}\varphi$,\label{axprob2}
				\item $\proba{\ge1}(\varphi\to\psi)\to(\proba{\ge p}\varphi\to\proba{\ge p}\psi)$.\label{axprob5}
			\end{enumerate}
		\end{enumerate}
		\begin{itemize}
			\item \underline{Necessitation Rule:} For the edge $S\overset{\ge1}{\rightsquigarrow}\Delta S$, if $\ded_S\varphi$, then $\ded_{\Delta S}\proba{\ge1}\varphi$.
			\item \underline{Archimedean Rule:} $\{\proba{\ge q}\psi\st q<p\}\ded_{\Delta S}\proba{\ge p}\psi$.
			\item \underline{Cover Rules:}
			\begin{enumerate}[\normalfont(1)]
				\item If $n\ge 1$, $\ded_{S}\varphi\to\bigvee_{I\subseteq J_m,\ \vabs{I}=n+k}\bigwedge_{i\in I}\varphi_i$, $\ded_{S}\bigvee_{I\subseteq J_m,\ \vabs{I}=k}\bigwedge_{i\in I}\varphi_i$, $p_1,\ldots,p_m\in [0,1]$, and $p=0\lor(1\land\frac{\sum_{i=1}^mp_i-k}{n})$, then
				\[\ded_{\Delta S}(\proba{\le p_1}\varphi_1\land\cdots\land\proba{\le p_m}\varphi_m)\to\proba{\le p}\varphi.\] \label{nkcoverruleprob1}
				\item If $\ded_S\bigvee_{I\subseteq J_m, \vabs{I}=k}\bigwedge_{i\in I}\varphi_i$, and $\sum_{i=1}^mp_i<k$ (with $ p_i\in [0,1]$), then \[\ded_{\Delta S}\neg(\proba{\le p_1}\varphi_1\land\cdots\land\proba{\le p_m}\varphi_m).\]\label{nkcoverruleprob2}
			\end{enumerate}
		\end{itemize}
	\end{defi}
	
	Axioms \ref{axprob}\ref{axprob1}-\ref{axprob5} are valid in all $T$-coalgebras. This is a consequence of Theorem \ref{soundness} and the fact that probability measures are upper probability measures. Also, the proof of the soundness of the inference rules for $\Delta S$ is as in Theorem \ref{conseqT}.
	
	The construction of the canonical coalgebra is easily adapted from the one presented in section \ref{seccanonicalspacesandcoalgebras}: if $\Delta S\in\ing T$, for each $\ded_{\Delta S}$-maximal set $x\in X_{\Delta S}^D$, define a function $\mu_x$ on $\Sigma_S^D$ by $\mu_x(\vabs{\varphi}_S)=\sup\{q\in[0,1]\st\proba{\ge q}\varphi\in x\}$.  Then we have:
	
	\begin{lem}\label{modalprinciplesfinitamenteaditivas}
		The following modal principles are valid in a $T$-deduction system:
		\begin{enumerate}[\normalfont(1)]
			\item $\ded_{\Delta S}\proba{\ge0}\varphi$.\label{propfadit1}
			\item If $p>0$, then $\ded_{\Delta  S}\proba{<p}\bot_{S}$,\label{propfadit2}
			\item If $\ded_S\varphi\leftrightarrow\psi$, then $\ded_{\Delta S}\proba{\ge p}\varphi\leftrightarrow\proba{\ge p}\psi$.\label{propfadit6}
			\item $\ded_{\Delta S}\proba{\ge q}\varphi\to\proba{\ge p}\varphi$ if $q>p$.\label{propfadit7}
			\item If $\ded_S\varphi$, then $\ded_{\Delta S}\proba{\ge p}\varphi$.\label{propfadit8}
		\end{enumerate}
	\end{lem}
	
	\begin{proof}
		The proof for items \ref{propfadit1}, \ref{propfadit6},\ref{propfadit7} and \ref{propfadit8} follows similar to Lemma \ref{dedproplema}. 
		
		The proof of item \ref{propfadit2} is easier because we are dealing now with a single modal operator: by axiom \ref{axprob}\ref{axprob1}, since $p>0$ we have that $\ded_{\Delta S}\proba{\le0}\bot_S\to\proba{<p}\bot_S$, or equivalently, $\ded_{\Delta S}\proba{\ge1}\top_S\to\proba{<p}\bot_S$. By the Necessitation rule, $\ded_{\Delta S}\proba{\ge 1}\top_S$, so by Detachment, the result follows.
	\end{proof}
	
	\begin{lem}\label{lema03082022}
		$\mu_x(\vabs{\varphi}_S)\ge p$ iff $\proba{\ge p}\varphi\in x$.
	\end{lem}
	
	\begin{proof} It is easy to check that the proof given in Lemma \ref{relacionmedcanocicaconformulas} \ref{23102021-1} works here as well.
	\end{proof}
	
	\begin{thm}\label{canonicalprobmeasurebuenadef}
		$\mu_x$ is a well defined probability measure on $\Sigma_{S}^D$.
	\end{thm}
	
	\begin{proof} First, we check $\mu_x$ is an upper probability measure. For \ref{30032024-1}, the proof is similar to the one given in Theorem \ref{canonicalmeasurebuenadef}, and using Lemma \ref{modalprinciplesfinitamenteaditivas} \ref{propfadit2}.
		
		For \ref{30032024-2} consider $\ded_S\top_S$. By the necessitation rule, it follows that $\ded_{\Delta S}\proba{\ge1}\top_S$ so $\proba{\ge1}\top_S\in x$. Therefore, $\mu_x(\vabs{\top_S}_S)=1$. \ref{30032024-3} is analogous to Theorem \ref{canonicalmeasurebuenadef}.
		
		For the additivity condition, it is enough to prove that $\mu_x(\vabs{\varphi}_S)=1-\mu_x(\vabs{\varphi}^c_S)$, for all $\vabs{\varphi}_S$, due to Observation \ref{whenupprobareprob}. The inequality $\mu_x(\vabs{\varphi}_S)\ge1-\mu_x(\vabs{\varphi}^c_S)$ holds because $\mu_x$ is an upper probability measure. If $\mu_x(\vabs{\varphi}_S)=p$, then we want to show that $p\le1-\mu_x(\vabs{\varphi}^c_S)$, which is equivalent to $p\le1-\mu_x(\vabs{\neg\varphi}_S)$.
		
		If it were the case that $p>1-\mu_x(\vabs{\neg\varphi}_S)$, then $1-p<\mu_x(\vabs{\neg\varphi}_S)$. Writing $q=\mu_x(\vabs{\neg\varphi}_S)$, and substituting $\varphi$ and $p$ for $\neg\varphi$ and $1-p$ respectively in axiom \ref{axprob}\ref{axprob1}, we obtain (i) $\ded_{\Delta S}\proba{\ge p}\varphi\to\neg\proba{\ge q}\neg\varphi$, since $1-p<q$.
		
		On the other hand, using Lemma \ref{lema03082022}, it follows that (ii) $\proba{\ge p}\varphi\in x$, and (iii) $\proba{\ge q}\neg\varphi\in x$.
		
		By (i), (ii) and Detachment, we have that $\neg\proba{\ge q}\neg\varphi\in x$, which contradicts (iii), so $\mu_x$ is finitely additive.
	\end{proof}

	\subsection{Plausibility measures and belief functions}
	
	In this section we expand our class of polynomial functors to include closure under the application of another variant of the functor $\Delta$, which we call $\Delta_{Pl}$. The aim is to provide another representation of uncertainty. We begin by outlining the history of the plausibility measures we are going to consider.
	
	In \cite{shafer76evidence}, Glenn Shafer called \textit{belief functions} those functions $\bbel:2^X\to [0,1]$ such that $\bbel(\emptyset)=0$, $\bbel(X)=1$, and for all $n\ge 1$ and $U_1,\ldots,U_n\subseteq X$ satisfy:
	\begin{equation}\label{shaferBel}
		\bbel(U_1\cup\ldots\cup U_n)\ge\displaystyle\sum_{\substack{I\subseteq J_n,\\ I\ne\emptyset}}(-1)^{\abs{I}+1}\bbel\left(\bigcap_{i\in I}U_i\right),
	\end{equation} where $X$ is a finite set. These functions were previously studied, in a wider context, by Choquet in \cite{choquet53capacities}, where he called them \textit{infinitely monotone capacities}, and also in the paper about upper and lower probabilities by Dempster \cite{dempster67upper}. In fact, in the finite case,  belief functions are a special case of lower probabilities \cite{halpern17reasoning}. Here we work with an algebra $\Sigma$ of subsets of a set $X$, not necessarily finite.
	
	The dual of a belief function $\bbel$ is a \emph{plausibility measure}, defined by $\pplaus(U)=1-\bbel(U^c)$, for all $U\in\Sigma$. Since these correspond in the finite case to upper measures, we take the plausibility measures as our primitives to mirror the work we did in the previous sections. 
	
	\begin{defi}\label{defplaus} 
		Given an uncertainty space $(X, \Sigma)$, a \emph{plausibility measure} over $X$ is a mapping
		$\plaus:\Sigma_X\to[0,1]$ such that $\plaus(\emptyset)=0$, $\plaus(X)=1$, and for all $U_1,\ldots,U_n\in\Sigma_X$ satisfies the following:
		\begin{equation}\label{12122021}
			\plaus(U_1\cap\ldots\cap U_n)\le\displaystyle\sum_{\substack{\emptyset\ne I\subseteq J_n}}(-1)^{\abs{I}+1}\plaus\left(\bigcup_{i\in I}U_i\right).
		\end{equation}
		We say that $\bel$ is a \emph{belief function} if for some plausibility measure $\plaus$, and all $U\in\Sigma_X$, $\bel(U)=1-\plaus(U^c)$. 
	\end{defi}
	
	\begin{lem}\label{lemapropbeliefyplaus}
		For any belief function $\bel$ and plausibility function $\plaus$ defined over an uncertainty space $(X,\Sigma)$, the following properties hold:
		\begin{enumerate}[\normalfont(1)]
			\item $\bel(\emptyset)=0$ and $\bel(X)=1$. \label{plandblpropemptyanduniverse}
			\item For all $U\in \Sigma$, $\bel(U)\le\plaus(U)$. \label{plandblorderwitheachother}
			\item \underline{Subaditivity:} if $U,V\in\Sigma$ are such that $U\cap V=\emptyset$, $\plaus(U\cup V)\le\plaus(U)+\plaus(V)$.\label{plandblpropsubaditivity}
			\item \underline{Superadditivity:} if $U,V\in\Sigma$ are such that $U\cap V=\emptyset$, $\bel(U\cup V)\ge\bel(U)+\bel(V)$.\label{plandblpropsuperaditivity}    
			\item \underline{Monotonicity:} if $U,V\in\Sigma$, and $U\subseteq V$, $\bel(U)\le\bel(V)$ and $\plaus(U)\le\plaus(V)$.\label{plandblpropmonotonocity}  
		\end{enumerate}
	\end{lem}
	
	\begin{proof}
		(1) follows directly from the definition. To prove \ref{plandblorderwitheachother}, let $U_1=U$ and $U_2=U^c$ in \ref{12122021} to obtain $0\le\plaus(U)+\plaus(U^c)-1$, so $\bel(U)=1-\plaus(U^c)\le\plaus(U)$.
		
		(3) follows from equation \ref{12122021} and (4) can easily be deduced from (3).
		
		For \ref{plandblpropmonotonocity}, suppose that $U\subseteq V$, then $\bel(V)=\bel(U\cup(V\cap U^c))\ge \bel(U)+\bel(V\cap U^c)\ge\bel(U)$. On the other hand, from $V^c\subseteq U^c$, it follows that $\bel(V^c)\le\bel(U^c)$, so $-\bel(U^c)\le-\bel(V^c)$ so $1-\bel(U^c)\le1-\bel(V^c)$, that is, $\plaus(U)\le\plaus(V)$.
	\end{proof}
	
	Now we can define for each uncertainty space $(X,\Sigma)$ the set $\Delta_{Pl}X$ of all the plausibility measures over $X$. Just as with $\Delta$, we can show that $\Delta_{Pl}X$ is a functor. To do this, we consider the algebra generated by the sets $\beta_{Pl}^{p,q}(U)=\{\plaus\in\Delta_{Pl}X\st\plaus(U)\ge p, 1-\plaus(U^c)\ge q\}$ for all $U\in \Sigma$, and we define the action of $\Delta_{Pl}$ on each measurable function $f$, by $(\Delta_{Pl}(f))(\plaus)(U)=\plaus(f^{-1}(U))$ for every $\plaus\in\Delta_{Pl}X$ and $U\in\Sigma$. We have the analogous to Lemma \ref{deltaesfuntorial}:
	
	\begin{lem}\label{deltaplesfuntorial}
		If $f:X\to Y$ is a  measurable function, defining  $\Delta_{Pl} f$  by $\Delta f(\plaus)(U)=\plaus(f^{-1}(U))$ for every $\plaus\in\Delta_{Pl} X$ and $U\in \Sigma_{Y}$, we have that 
		$(\Delta f)(\plaus)$ is a plausibility measure and  $\Delta_{Pl} f: \Delta X\to \Delta Y$ is a measurable function. As a consequence, $\Delta_{Pl}$ is an endofunctor.
	\end{lem}
	
	\begin{proof} We need to prove that for every family of sets $U_1,\ldots, U_n\in \Sigma_X$,
		$(\Delta_{Pl}f)(\plaus)(U_1\cap\ldots\cap U_n)\le\sum_{I\subseteq J_n,I\ne\emptyset}(-1)^{\abs{I}+1}(\Delta_{Pl}f)(\plaus)(\bigcup_{i\in I}U_i)$, but this is equivalent to $\plaus(f^{-1}(U_1)\cap\ldots\cap f^{-1}(U_n))\le\sum_{I\subseteq J_n,I\ne\emptyset}(-1)^{\abs{I}+1}\plaus(\bigcup_{i\in I}f^{-1}(U_i))$ $=$
		$\sum_{I\subseteq J_n,I\ne\emptyset}(-1)^{\abs{I}+1}\plaus(f^{-1}(\bigcup_{i\in I}(U_i))$, which is true because $\plaus$ is a plausibility function. The rest of the proof is as in Lemma \ref{losbetasmedibles}.
	\end{proof}
	
	We expand the definition of our polynomial uncertainty functors to include closure under the application of $\Delta_{Pl}$, and for each polynomial uncertainty functor, the multigraph of ingredients will have edges from $\Delta_{Pl}S$  to $S$ labeled with $(p,q)_{Pl}$, for each $p,q\in[0,1]$.
	
	\begin{defi}
		For every  formula $\varphi::S$, $[(p,q)]_{Pl}\varphi$ is a formula of type $\Delta_{Pl}S$, with semantics $\class{[(p,q)]_{Pl}\varphi}_{\Delta_{Pl}S}^\alpha=\beta_{Pl}^{p,q}\class{\varphi}_S^\alpha$.
		The associated modal operators $[(p,q)]_{Pl}$ will be abbreviated in the following way: 
		\begin{itemize}
			\item $\pl{\ge p}\varphi$ stands for $[(p,0)]_{Pl}\varphi$. Thus $\plaus\in\class{\pl{\ge p}\varphi}_{\Delta_{Pl}S}^\alpha$ iff $\plaus(\class{\varphi}_{S}^\alpha)\ge p$. Since this is equivalent to $1-\plaus(SX-\class{\neg\varphi}_{S}^\alpha)\le 1-p$, we also write $[(p,0)]_{Pl}\varphi$ as $\bl{\le 1-p}\neg\varphi$.
			\item $\bl{\ge p}\varphi$ and $\pl{\le 1-p}\neg\varphi$  both stand for $[(0,p)]_{Pl}\varphi$, so  $\plaus\in\class{\bl{\ge p}\varphi}_{\Delta_{Pl}S}^\alpha$ iff $1-\plaus(SX-\class{\varphi}_{S}^\alpha)\ge p$.
		\end{itemize}

		Similarly, we also write  $\pl{\le p}\varphi$ or $\bl{\ge 1-p}\neg\varphi$ for $[(0,1-p)]_{Pl}\neg\varphi$;
		$\bl{\le p}\varphi$ or $\pl{\ge 1-p}\neg\varphi$  for $[(1-p,0)]_{Pl}\neg\varphi$;
		$\neg\pl{\ge p}\varphi$ or $\pl{<p}\varphi$ for $\neg[(p,0)]_{Pl}\varphi$, and 
		$\neg\bl{\ge p}\varphi$ or $\bl{<p}\varphi$  for $\neg[(0,p)]_{Pl}\varphi$.
		
		With $\pl{[i]p}\varphi$ we will denote the  modal formulas:
		\[
		\pl{[i]p}\varphi=\left\{\begin{array}{ll}
			\pl{\le p}\varphi & \text{if } i \text{ is odd} \\
			\\
			\pl{\ge p}\varphi & \text{if } i \text{ is even}.
		\end{array}
		\right.
		\] Furthermore, if $\varphi_1,\ldots ,\varphi_n$ are formulas of the same type, and $I$ is a nonempty subset of $J_n$, we denote with $\psi_I$ the disjunction $\bigvee_{i\in I}\varphi_i$.
	\end{defi}
	We now list the corresponding axioms and inference rules:
	\begin{defi}\label{plausAxiomsAndRules}
		\begin{enumerate}[\normalfont(1)]\setcounter{enumi}{8}
			\item For an ingredient of the form $\Delta_{Pl}S$,\label{axplaus}
			\begin{enumerate}[\normalfont(a)]
				\item $\pl{\le p}\varphi\to\pl{<q}\varphi$ for $p<q$,\label{axplaus1}
				\item $\pl{<p}\varphi\to\pl{\le p}\varphi$,\label{axplaus2}
				\item $\pl{\le p}\varphi\to\bl{\le p}\varphi$,\label{axplaus3}
				\item $\bl{\ge1}(\varphi\to\psi)\to(\pl{\ge p}\varphi\to \pl{\ge p}\psi)$,\label{axplaus4}
				\item $[(p,q)]_{pl}\varphi\leftrightarrow(\pl{\ge p}\varphi\land \bl{\ge q}\varphi)$,\label{axplaus5}
			\end{enumerate}
			For each nonempty set $I\subseteq J_n$, let $p_I\in[0,1]$, and $q=\sum_{I\subseteq J_n}(-1)^{|I|+1}p_I$.
			\begin{enumerate}[\normalfont(a)]\setcounter{enumii}{5}
				\item If $q\ge0$, and $p=q\land 1$, then the following is an axiom:
				\[\displaystyle\left(\bigwedge_{I\subseteq J_n}\pl{[|I|]p_I}\psi_I\right)\to\pl{\le p}\left(\bigwedge_{i=1}^n\varphi_i\right).
				\]\label{axplaus6}
				\item If $q<0$, then we have the axiom
				\[\displaystyle\neg\left(\bigwedge_{I\subseteq J_n}\pl{[|I|]p_I}\psi_I\right).
				\]\label{axplaus7}
			\end{enumerate}
		\end{enumerate}
		
		\begin{itemize}   \label{defdeductivesystemplaus}
			\item \underline{Necessitation Rule:} For the edge $S\overset{(0,1)_{Pl}}{\rightsquigarrow}\Delta_{Pl}S$, if $\ded_S\varphi$, then $\ded_{\Delta_{Pl}S}\bl{\ge1}\varphi$.
			\item \underline{Archimedean Rule:} $\{\pl{\ge q}\psi\st q<p\}\ded_{\Delta_{Pl}S}\pl{\ge p}\psi$, and $\{\bl{\ge q}\psi\st q<p\}\ded_{\Delta_{Pl}S}\bl{\ge p}\psi$.
		\end{itemize}
	\end{defi}
	
	\begin{lem}\label{modalprinciplesplaus}
		The following modal principles are valid in a $T$-deduction system:
		\begin{enumerate}[\normalfont(1)]
			\item $\ded_{\Delta_{Pl}S}\pl{\ge0}\varphi$ and $\ded_{\Delta_{Pl}S}\bl{\ge0}\varphi$.\label{proppl1}
			\item If $p>0$, then $\ded_{\Delta_{Pl} S}\pl{<p}\bot_{S}$, and $\ded_{\Delta_{Pl}S}\bl{<p}\bot_S$.\label{proppl2}
			\item If $\ded_S\varphi\leftrightarrow\psi$, then $\ded_{\Delta_{Pl}S}\pl{\ge p}\varphi\leftrightarrow\pl{\ge p}\psi$.\label{proppl6}
			\item $\ded_{\Delta_{Pl}S}\pl{\ge q}\varphi\to\pl{\ge p}\varphi$ if $q>p$.\label{proppl7}
			\item If $\ded_S\varphi$, then $\ded_{\Delta_{Pl}S}\pl{\ge p}\varphi$.\label{proppl8}
		\end{enumerate}
	\end{lem}
	
	\begin{proof}
		We prove item \ref{proppl2} only, since the rest of the proofs are analogous to those of Lemma \ref{dedproplema}. Since $p>0$, by \ref{axplaus}\ref{axplaus1}, we have that $\ded_{\Delta_{Pl}}\pl{\le0}\bot_S\to\pl{<p}\bot_S$, which is equivalent to  $\ded_{\Delta_{Pl}}\bl{\ge1}\top_S\to\pl{<p}\bot_S$. Using the necessitation rule we also have that $\ded_{\Delta_{Pl}}\bl{\ge1}\top_S$, so $\ded_{\Delta_{Pl}}\pl{<p}\bot_S$.
		
		By axiom \ref{axplaus}\ref{axplaus3}, we have that for any $\varphi::S$, $\ded_{\Delta_{Pl}S}\pl{\le 1-p}\neg\varphi\to\bl{\le 1-p}\neg\varphi$, which we can write as  $\ded_{\Delta_{Pl}S}\bl{\ge p}\varphi\to\pl{\ge p}\varphi$.  By contraposition, we have that $\ded_{\Delta_{Pl}S}\neg\pl{\ge p}\bot_S\to\neg\bl{\ge p}\bot_S$, so $\ded_{\Delta_{Pl}S}\pl{<p}\bot_S\to\bl{<p}\bot_S$, and thus we get $\ded_{\Delta_{Pl}S}\bl{<p}\bot_S$.
	\end{proof}
	
	\begin{lem}\label{plaussoundness}
		For any $S\in\ing T$, all axioms are valid in all $T$-coalgebras, and the inference rules for the ingredient $\Delta_{Pl}$ also preserve validity.
	\end{lem}
	
	\begin{proof}
		We only need to show the validity of the axioms \ref{axplaus}\ref{axplaus1} to  \ref{axplaus}\ref{axplaus6}. Items \ref{axplaus1} and \ref{axplaus2} follow from the fact plausibility measures are real-valued functions, and the reals are linearly ordered. Item \ref{axplaus3} is a consequence of Lemma \ref{lemapropbeliefyplaus} \ref{plandblorderwitheachother} the fact that for all $U\in\Sigma$, $\bel(U)\le\plaus(U)$. The proof of \ref{axplaus}\ref{axplaus4} is similar to the proof for axiom \ref{axupprob}\ref{axupprob5}, since the subadditivity and monotonicity properties hold for plausibility measures.
		
		Let us check the validity of \ref{axplaus}\ref{axplaus6}. Given a $T$-coalgebra $(X,\alpha)$ such that $\deltapl S\in\ing(T)$, and  $\plaus\in\deltapl SX$, assume that for each nonempty $I\subseteq J_n$, $\alpha, \plaus\sat^\alpha_{\Delta_{Pl} S}\pl{[\vabs{I}]p_j}\psi_I$.
		
		If $\vabs{I}$ is odd, $\alpha,\plaus\sat_{\Delta_{Pl} S}\pl{\le p_I}\psi_I$, so by the semantics it follows that (i) $\plaus(\class{\psi_I})=\plaus\left(\bigcup_{i\in I}\class{\varphi_i}^\alpha_{S}\right)\le p_I$. Analogously, if $\vabs{I}$ is even, we conclude that $\alpha,\plaus\sat_{\Delta_{Pl} S}\pl{\ge p_I}\psi_I$ so (ii) $-\plaus\left(\bigcup_{i\in I}\class{\varphi_i}^\alpha_{S}\right)\le -p_I$. Adding (i) and (ii)  over all the nonempty subsets of $J_n$ we obtain
		\[
		\sum_{I\subseteq J_n}(-1)^{\vabs{I}+1}\plaus\left(\bigcup_{i\in I}\class{\varphi_i}^\alpha_S\right)\le\sum_{I\subseteq J_n}(-1)^{\vabs{I}+1}p_I=q.
		\]
		Since $\plaus$ is a plausibility measure, \ref{12122021} gives $0\le \plaus\left(\bigcap_{i=1}^n\class{\varphi_i}^\alpha_S\right)\le q$, so for $p=q\land 1$ we have that $\alpha,\plaus\sat_{\Delta_{Pl} S}\pl{\le p}\left(\bigwedge_{i=1}^n\varphi_i\right)$, hence \ref{axplaus}\ref{axplaus6} is valid. 
		
		To prove the validity of axiom \ref{axplaus}\ref{axplaus7}, assume towards contradiction that \linebreak[4] $q=$ $\sum_{I\subseteq J_n}(-1)^{|I|+1}p_I$ $<0$ and that there exists a plausibility measure $\plaus$ such that $\alpha,\plaus\not\sat_{\Delta_{Pl} S}\neg(\bigwedge_{I\subseteq J_n}\pl{[|I|]p_I}\psi_I)$. Then $\alpha,\plaus\sat_{\Delta_{Pl} S}(\bigwedge_{I\subseteq J_n}\pl{[|I|]p_I}\psi_I)$, so $\alpha,\plaus\sat_{\Delta_{Pl} S}\pl{[|I|]p_I}\psi_I$ for all nonempty $I\subseteq J_n$. By the semantics, it follows that $\plaus(\class{\bigvee_{i\in I}\varphi_i})\le p_I$, for $\vabs{I}$ odd, and $\plaus(\class{\bigvee_{i\in I}\varphi_i})\ge p_I$, for $\vabs{I}$ even. By \ref{12122021}, it follows that $0\le\plaus(\bigcap_{i=1}^n\class{\varphi_i})\le\sum_{\substack{\emptyset\ne I\subseteq J_n}}(-1)^{\abs{I}+1}\plaus\left(\bigcup_{i\in I}\class{\varphi_i}\right)\le\sum_{\substack{\emptyset\ne I\subseteq J_n}}(-1)^{\abs{I}+1}p_I=q<0$, which is a contradiction.
		
		The proof that the inference rules preserve validity follows as in Theorem \ref{conseqT}.
	\end{proof}
	
	We construct once again the canonical measure as expected, by defining for $\Delta_{Pl}S\in\ing T$, each $\ded_{\Delta^*S}$-maximal set $x\in X_{\Delta_{Pl}S}^D$, and $\varphi::S$, $\plaus_x(\vabs{\varphi}_S)=\sup\{q\in[0,1]\st\pl{\ge q}\varphi\in x\}$. Using axiom \ref{axplaus}\ref{axplaus4}, it is easy to prove that $\plaus_x$ is monotone.
	
	\begin{lem}\label{propmedcanplaus}
		For each $x\in X_{\Delta_{Pl}}^D$
		\begin{enumerate}[\normalfont(1)]
			\item $\plaus_x(\vabs{\varphi}_S)<p$ implies $\neg\pl{\ge p}\varphi\in x$.\label{propmedcanplaus1}
			\item $\plaus_x(\vabs{\varphi}_S)>p$ implies $\pl{\le p}\varphi\notin x$.\label{propmedcanplaus2}
			\item $\plaus_x(\abs{\varphi})\ge p$ implies $\pl{\ge p}\varphi\in x$. \label{propmedcanplaus3} 
		\end{enumerate}
	\end{lem}
	
	\begin{proof}
		Items \ref{propmedcanplaus1} and \ref{propmedcanplaus2} are proved as in Lemma \ref{propmedcan}. The proof of item \ref{propmedcanplaus3} is analogous to the proof of Lemma \ref{relacionmedcanocicaconformulas}\ref{23102021-1}, and uses Lemma \ref{modalprinciplesplaus}\ref{proppl7}.
	\end{proof}
	\begin{thm}\label{canonicalplausmeasurebuenadef}
		$\plaus_x$ is a well-defined plausibility measure on $\Sigma_S^D$.
	\end{thm}
	
	\begin{proof}
		Reasoning along the lines of Theorem \ref{canonicalmeasurebuenadef}, it follows that the definition does not depend on the formulas,  $\plaus_x(\emptyset)=0$ (using Lemma \ref{modalprinciplesplaus}\ref{proppl2}), and $\plaus_x(X)=1$. It remains to show that $\plaus_x$ satisfies condition \ref{12122021}, that is $\plaus_x(\vabs{\varphi_1\land\ldots\land\varphi_n}_S)\le\sum_{\substack{\emptyset\ne I\subseteq J_n}}(-1)^{\abs{I}+1}\plaus_x\left(\vabs{\psi_I}_S\right)$. For $n=1$ this is trivial.
		
		For $n\ge2$, assume towards contradiction that $A=\sum_{\substack{\emptyset\ne I\subseteq J_n}}(-1)^{\abs{I}+1}\plaus_x\left(\vabs{\psi_I}_S\right)$ $<$\linebreak[4] $\plaus_x(\vabs{\varphi_1\land\ldots\land\varphi_n}_S)$ $=B$, we have two cases to consider:
		
		(I) $A<0$. Then, for each nonempty subset $I$ of $J_n$, we can find positive real numbers $\varepsilon_I$ small enough so that if we define
		\[p_I=\left\{\begin{array}{lc}\plaus_x(\vabs{\psi_I}_S)+\varepsilon_I & \text{if }\vabs{I}\text{ is odd and }\plaus_x(\vabs{\psi_I}_S)\ne 1 \\ \plaus_x(\vabs{\psi_I}_S)-\varepsilon_I & \text{if }\vabs{I}\text{ is even and }\plaus_x(\vabs{\psi_I}_S)\ne 0 \\ \plaus_x(\vabs{\psi_I}_S) & \text{ otherwise}\end{array}\right.\]
		we have that  $p_I\in[0,1]$ and also
		\begin{equation}\label{31072023}
			\sum_{\substack{\emptyset\ne I\subseteq J_n}}(-1)^{\abs{I}+1}\plaus_x(\vabs{\psi_I}_S)\le\sum_{\substack{\emptyset\ne I\subseteq J_n}}(-1)^{\abs{I}+1}p_I<0.
		\end{equation}
		
		In order to prove the existence of these $p_I$, let $\varepsilon=-\frac{A}{2^n}>0$. Then we take 
		\[\varepsilon_I=\left\{\begin{array}{lc}\min\{\varepsilon,(1-\plaus_x(\vabs{\psi_I}_S))/2\} & \text{if } \vabs{I}\text{ is odd and }\plaus_x(\vabs{\psi_I}_S)\ne1 \\ 
			\min\{\varepsilon,\plaus_x(\vabs{\psi_I}_S))/2\} & \text{if } \vabs{I}\text{ is even and }\plaus_x(\vabs{\psi_I}_S)\ne0.\end{array} \right.\]
		Defining $p_I$ as above, and letting $q=\sum_{\substack{\emptyset\ne I\subseteq J_n}}(-1)^{\abs{I}+1}p_I$ we have that
		\[
		A\le q\le\displaystyle\sum_{\substack{\emptyset\ne I\subseteq J_n\\ \vabs{I}\text{ odd}}}(\plaus_x(\vabs{\psi_I}_S)+\varepsilon)-\displaystyle\sum_{\substack{\emptyset\ne I\subseteq J_n\\\vabs{I}\text{ even}}}(\plaus_x(\vabs{\psi_I}_S)-\varepsilon)=A+(2^n-1)\varepsilon=
		\]
		\[
		= A+(2^n-1)\varepsilon+\varepsilon-\varepsilon = A+2^n\varepsilon-\varepsilon=A-A-\varepsilon<0.
		\]
		
		Since $q<0$, axiom \ref{axplaus}\ref{axplaus6} gives $\neg(\bigwedge_{I\subseteq J_n}\pl{[|I|]p_I}\psi_I)\in x$, so $\bigvee_{I\subseteq J_n}\neg\pl{[|I|]p_I}\psi_I\in x$, and therefore there exists $\emptyset\ne I_0\subseteq J$ such that (i) $\neg\pl{[|I|]p_{I_0}}\psi_{I_0}\in x$. We have two cases: 
		\begin{itemize}
			\item If $\vabs{I_0}$ is odd, then by (i) $\neg\pl{\le p_{I_0}}\psi_{I_0}\in x$. We have that $p_{I_0}<1$. Indeed, if it were the case that $p_{I_0}=1$, then $\neg\pl{\le 1}\psi_{I_0}\in x$, and by Lemma \ref{modalprinciplesplaus} \ref{proppl1}, $\bl{\ge 0}\neg\psi_{I_0}$, which is equivalent to $\pl{\le 1}\psi_{I_0}\in x$, which contradicts the consistency of $x$. From the definition of $p_I$, we had that in this case (ii) $\plaus_x(\vabs{\psi_{I_0}}_S)<p_{I_0}$. From (i) we deduce that $\pl{>p_{I_0}}\psi_{I_0}\in x$, and using axiom \ref{axplaus}\ref{axplaus2} it follows that $\pl{\ge p_{I_0}}\psi_{I_0}\in x$. By Lemma \ref{propmedcanplaus} \ref{propmedcanplaus1}, we have that $\plaus_x(\vabs{\psi_{I_0}}_S)\ge p_{I_0}$, which contradicts (ii).
			\item If $\vabs{I_0}$ is even, then by (i) $\neg\pl{\ge p_{I_0}}\psi_{I_0}\in x$. We have that $p_{I_0}>0$. Indeed, if it were the case that $p_{I_0}=0$, then $\neg\pl{\ge 0}\psi_{I_0}\in x$, and by Lemma \ref{modalprinciplesplaus} \ref{proppl1}, $\pl{\ge 0}\psi_{I_0}\in x$, which is impossible. Therefore, by the definition of $p_{I_0}$, (iii) $\plaus_x(\vabs{\psi_{I_0}}_S)>p_{I_0}$. On the other hand, we had  $\pl{<p_{I_0}}\psi_{I_0}\in x$, and using axiom \ref{axplaus}\ref{axplaus2} it follows that $\pl{\le p_{I_0}}\psi_{I_0}\in x$. By Lemma \ref{propmedcanplaus} \ref{propmedcanplaus2}, we have that $\plaus_x(\vabs{\psi_I}_S)\le p_{I_0}$, which contradicts (iii).
		\end{itemize}
		
		(II) Now we assume that 
		\[0\le A=\sum_{\substack{\emptyset\ne I\subseteq J_n}}(-1)^{\abs{I}+1}\plaus_x\left(\vabs{\psi_I}_S\right)<\plaus_x(\vabs{\varphi_1\land\ldots\land\varphi_n}_S)=B\]
		
		Let $\varepsilon=\frac{B-A}{2^n}$, and let as before, for each nonempty subset $I$  of $J_n$, 
		\[\varepsilon_I=\left\{\begin{array}{lc}\min\{\varepsilon,(1-\plaus_x(\vabs{\psi_I}_S))/2\} & \text{if } \vabs{I}\text{ is odd and }\plaus_x(\vabs{\psi_I}_S)\ne1 \\ 
			\min\{\varepsilon,\plaus_x(\vabs{\psi_I}_S))/2\} & \text{if } \vabs{I}\text{ is even and }\plaus_x(\vabs{\psi_I}_S)\ne0.\end{array} \right.\]
		
		Now define
		\[p_I=\left\{\begin{array}{lc}\plaus_x(\vabs{\psi_I})+\varepsilon_I & \text{if }\vabs{I}\text{ is odd and }\plaus_x(\vabs{\psi_I})\ne 1 \\ \plaus_x(\vabs{\psi_I})-\varepsilon_I & \text{if }\vabs{I}\text{ is even and }\plaus_x(\vabs{\psi_I})\ne 0 \\ \plaus_x(\vabs{\psi_I}) & \text{ otherwise.}\end{array}\right.\]

		Then we have that $p_I\in[0,1]$ and 
		\[ A\le \sum_{\emptyset\ne I\subseteq J_n}(-1)^{\abs{I}+1}p_I  \le \sum_{\emptyset\ne I\subseteq J_n}(\plaus_x(\vabs{\psi_I}_S)+\varepsilon)-\sum_{\emptyset\ne I\subseteq J_n}(\plaus_x(\vabs{\psi_I}_S)-\varepsilon)
		=
		\]
		\[A+(2^n-1)\varepsilon=A+(2^n-1)\varepsilon+\varepsilon-\varepsilon=A+2^n\varepsilon-\varepsilon=A+{B-A}-\varepsilon<B=\plaus_x(\vabs{\varphi_1\land\ldots\land\varphi_n}_S).
		\]
		So in particular we have:
		\begin{equation}\label{28042023}
			\displaystyle\sum_{\substack{\emptyset\ne I\subseteq J_n}}(-1)^{\abs{I}+1}\plaus_x\left(\vabs{\psi_I}_S\right)\le \displaystyle\sum_{\substack{\emptyset\ne I\subseteq J_n}}(-1)^{\abs{I}+1}p_I<\plaus_x(\vabs{\varphi_1\land\ldots\land\varphi_n}_S)
		\end{equation}
		
		If $\vabs{I}$ is odd and $\plaus_x(\vabs{\psi_I})= 1$, we have that $p_I=1$ and by Lemma \ref{modalprinciplesplaus} \ref{proppl1},  $\bl{\ge 0}\neg\psi_{I}\in x$, which is equivalent to  (iv) $\pl{\le 1}\psi_{I}\in x$. If $\plaus_x(\vabs{\psi_I})<p_I<1$, then by Lemma \ref{propmedcanplaus}\ref{propmedcanplaus1},  $\neg\pl{\ge p_I}\psi_I\in x$. This is equivalent to $\pl{<p_I}\psi_I\in x$. By axiom \ref{axplaus}\ref{axplaus2} and detachment we have that (v) $\pl{\le p_I}\psi_I\in x$. 
		
		If $\vabs{I}$ is even, from  $\plaus_x(\vabs{\psi_I})\ge p_I$, and Lemma \ref{propmedcanplaus}\ref{propmedcanplaus3} it follows that (vi) $\pl{\ge p_I}\psi_I\in x$.
		
		From (iv), (v) and (vi), it follows that $\bigwedge_{I\subseteq J_n}\pl{[|I|]p_I}\psi_I\in x$, so by axiom \ref{axplaus}\ref{axplaus6}, $\pl{\le p}(\bigwedge_{i=1}^n\varphi_i)\in x$, where $p=1\land\sum_{\emptyset\ne I\subseteq J_n}(-1)^{\abs{I}+1}p_I$, so by the contrapositive of Lemma \ref{propmedcanplaus}\ref{propmedcanplaus2}, we have that $\plaus_x(\vabs{\bigwedge_{i=1}^n\varphi_i}_S)\le p$, which contradicts \ref{28042023}.
	\end{proof}

	\subsection{Possibility and Necessity measures}
	
	\begin{defi}\label{defposs}
		Given an uncertainty space $(X, \Sigma_X)$, a \emph{possibility measure} is a function $\poss:\Sigma_X\to[0,1]$, such that the following conditions are verified:
		\begin{enumerate}[\normalfont(Poss 1)]
			\item $\poss(\emptyset)=0$,\label{06072023-1}
			\item $\poss(X)=1$,\label{06072023-2}
			\item If $U,V\in\Sigma$, then $\poss(U\cup V)=\max\{\poss(U),\poss(V)\}$.\label{06072023-3}
		\end{enumerate}
	\end{defi}
	
	It can be proved that possibility measures are plausibility measures, and therefore, in the finite case they are also upper probabilities. The dual of possibility measures are \emph{necessity measures}, $\nec:\Sigma_X\to[0,1]$, defined for all $U\in\Sigma_X$ as $\nec(U)=1-\poss(U^c)$.
	
	\begin{lem}\label{lemaproppossyness}
		For any possibility measure $\poss$ and necessity measure $\nec$ defined over an uncertainty space $(X,\Sigma)$, the following properties hold:
		\begin{enumerate}[\normalfont(1)]
			\item For all $U\in \Sigma$, $\nec(U)\le\poss(U)$. \label{psandncorderwitheachother}
			\item \underline{Monotonicity:} if $U,V\in\Sigma$, and $U\subseteq V$, $\poss(U)\le\poss(V)$ and $\nec(U)\le\nec(V)$.\label{psandncpropmonotonocity}
		\end{enumerate}
	\end{lem}
	
	\begin{proof}
		\ref{psandncorderwitheachother} From $1=\poss(U\cup U^c)=\max\set{\poss(U),\poss(U^c)}$ it follows that $\poss(U)=1$ or $\poss(U^c)=1$. In any case, $\poss(U)+\poss(U^c)\ge 1$, so $\nec(U)=1-\poss(U^c)\le\poss(U)$.
		
		\ref{psandncpropmonotonocity} If $U\subseteq V$, then $\poss(U)\le \max\set{\poss(U),\poss(v)}=\poss(U\cup V)=\poss(V)$. The other inequality follows from definition of $\nec$.
	\end{proof}
	
	Now we can define for each uncertainty space $(X,\Sigma)$ the set $\Delta_{Ps}X$ of all the possibility measures over $X$. We need to show that $\Delta_{Ps}$ is a functor. To do this, we consider the algebra generated by the sets $\beta_{Ps}^{p,q}(U)=\{\poss\in\Delta_{Ps}X\st\poss(U)\ge p, 1-\poss(U^c)\ge q\}$ for all $U\in \Sigma$, and we define the action of $\Delta_{Ps}$ on each measurable function $f$, by $(\Delta_{Ps}(f))(\poss)(U)=\poss(f^{-1}(U))$ for every $\poss\in\Delta_{Ps}X$ and $U\in\Sigma$. This yields a measurable function:
	
	\begin{lem}\label{deltapsesfuntorial}
		Let $X$ and $Y$ be uncertainty spaces, and $f:X\to Y$ a measurable function. If $\poss\in \Delta_{Ps}(X)$, then $(\Delta_{Ps}f)(\poss)\in \Delta_{Ps}Y$.
	\end{lem}
	
	\begin{proof} 
		\ref{06072023-1} and \ref{06072023-2} follow as \ref{30032024-1} and \ref{30032024-2} in Lemma \ref{caracupprob}. 
			
			To see \ref{06072023-3}, let $U,V\in\Sigma_X$. Then $(\Delta_{Ps}(f))(\poss)(U\cup V)=$ $\poss(f^{-1}(U\cup V))=\poss(f^{-1}(U)\cup f^{-1}(V))=$ $\max\{\poss(f^{-1}(U)),$ $\poss(f^{-1}(V))\}=\max\{(\Delta_{Ps}(f))(\poss)(U),(\Delta_{Ps}(f))(\poss)(V)\}$. Therefore, $(\Delta_{Ps}(f))(\poss)$ is a possibility measure. 
\end{proof}
	
	With the new ingredients of the form $\Delta_{Ps}S$ come  modal formulas: if $\varphi::S$, then  the formula $[(p,q)]_{Ps}\varphi$ is of sort $\Delta_{Ps}S$, and has semantics $\class{[(p,q)]_{Ps}\varphi}^\alpha_{\Delta_{Ps}S}=\beta_{Ps}^{p,q}(\class{\varphi}^\alpha_S)$. With $\pposs{\ge p}\varphi$ and $\nnec{\ge p}\varphi$ we abbreviate $[(p,0)]_{Ps}\varphi$ and $[(0,q)]_{Ps}\varphi$ respectively.
	
	We now give the corresponding list of axioms and inference rules:
	
	\begin{defi}\label{possAxiomsAndRules}
		\begin{enumerate}[\normalfont\ (1)]\setcounter{enumi}{9}
			\item For an ingredient of the form $\Delta_{Ps}S$,\label{axposs}
			\begin{enumerate}[\normalfont(a)]
				\item $\pposs{\le p}\varphi\to\pposs{<q}\varphi$ for $p<q$,\label{axposs1}
				\item $\pposs{<p}\varphi\to\pposs{\le p}\varphi$,\label{axposs2}
				\item $\pposs{\le0}\bot_S$, \label{axposs3}
				\item $\pposs{\le p}\varphi\to\nnec{\le p}\varphi$, \label{axposs4} 
				\item $\pposs{\le p}\varphi\land\pposs{\le q}\psi\to\pposs{\le p\lor q}(\varphi\lor\psi)$, \label{axposs5}
				\item $\nnec{\ge1}(\varphi\to\psi)\to(\pposs{\ge p}\varphi\to \pposs{\ge p}\psi)$,\label{axposs6}
				\item $[(p,q)]_{Ps}\varphi\leftrightarrow(\pposs{\ge p}\varphi\land\nnec{\ge q}\varphi)$,\label{axposs7}
			\end{enumerate}
		\end{enumerate}
		\begin{itemize}\label{defdeductivesystemposs}
			\item \underline{Necessitation Rule:} For the edge $S\overset{(0,1)_{Ps}}{\rightsquigarrow}\Delta_{Ps}S$, if $\ded_S\varphi$, then $\ded_{\Delta_{Ps}S}\nnec{\ge1}\varphi$.
			\item \underline{Archimedean Rule:} $\{\pposs{\ge q}\psi\st q<p\}\ded_{\Delta_{Pl}S}\pposs{\ge p}\psi$, and $\{\nnec{\ge q}\psi\st q<p\}\ded_{\Delta_{Pl}S}\nnec{\ge p}\psi$.
		\end{itemize}
	\end{defi}
	
	\begin{lem}\label{modalprinciplesposs}
		The following modal principles are valid in a $T$-deduction system:
		\begin{enumerate}[\normalfont(1)]
			\item $\ded_{\Delta_{Ps}S}\pposs{\ge0}\varphi$ and $\ded_{\Delta_{Ps}S}\nnec{\ge0}\varphi$.\label{propps0}
			\item $\ded_{\Delta_{Ps}S}\nnec{\ge p}\varphi\to\pposs{\ge p}\varphi$.\label{propps1}
			\item If $p>0$, then $\ded_{\Delta_{Ps} S}\pposs{<p}\bot_{S}$.\label{propps2}
			\item If $\ded_S\varphi\leftrightarrow\psi$, then $\ded_{\Delta_{Ps}S}\pposs{\ge p}\varphi\leftrightarrow\pposs{\ge p}\psi$.\label{propps3}
			\item $\ded_{\Delta_{Ps}S}\pposs{\ge q}\varphi\to\pposs{\ge p}\varphi$ if $q>p$.\label{propps4}
			\item If $\ded_S\varphi$, then $\ded_{\Delta_{Ps}S}\pposs{\ge p}\varphi$.\label{propps5}
		\end{enumerate}
	\end{lem}
	
	\begin{proof} \ref{propps0} Can be obtained using the Archimedean rule as in Lemma \ref{dedproplema}\ref{prop1}.
		
		\ref{propps1} By axiom \ref{axposs}\ref{axposs4} $\ded_{\Delta_{Ps}S}\pposs{\le 1-p}\neg\varphi\to\nnec{\le 1-p}\neg\varphi$, which is equivalent to $\ded_{\Delta_{Ps}S}\nnec{\ge p}\varphi\to\pposs{\ge p}\varphi$.
		
		Item \ref{propps2} follows from axioms \ref{axposs}\ref{axposs3} and \ref{axposs}\ref{axposs1}, while \ref{propps3} follows directly from axiom \ref{axposs}\ref{axposs6}. \ref{propps4} Is proved as in Lemma \ref{dedproplema} \ref{prop7}.
		
		\ref{propps5} By the necessitation rule, item \ref{propps1} and the hypothesis, $\ded_{\Delta_{Ps}S}\pposs{\ge 1}\varphi$. If $p<1$, using \ref{propps4} we have that $\ded_{\Delta_{Ps}S}\pposs{\ge p}\varphi$.
	\end{proof}
	
	\begin{lem}\label{possoundness}
		For any functor $T$ having an ingredient of the form $\Delta_{Ps}S$, axioms \ref{axposs}\ref{axposs1} to \ref{axposs}\ref{axposs7} are valid in all $T$-coalgebras, and the corresponding inference rules preserve validity as well.
	\end{lem}
	
	\begin{proof}
		The validity of axioms \ref{axposs1} and \ref{axposs2} follow from the facts that $\poss(U)\le p$ implies $\poss(U)<q$ if $p<q$, and $\poss(U)<p$ implies $\poss(U)\le p$. Item \ref{axplaus3} is a consequence of the property \ref{06072023-1}, while item \ref{axplaus4} can be deduced from the property $\nec(U)\le\poss(U)$, for all $U\in\Sigma$.
		
		Axiom \ref{axposs}\ref{axposs5} follows from \ref{06072023-3}. To prove  axiom \ref{axposs}\ref{axposs6}, assume that $\nec(\class{\varphi\to\psi})=1$ so $1-\poss(\class{\neg\varphi\lor\psi}^c)=1$ and therefore (i) $\poss(\class{\varphi\land\neg\psi})=0$. We also assume that $\poss(\class{\varphi})\ge p$. Since $\poss(\class{\varphi})=\poss(\class{\varphi\land\neg\psi}\cup \class{\varphi\land\psi}))=\max\set{\poss(\class{\varphi\land\neg\psi}),\poss( \class{\varphi\land\psi}))}$, from (i) we get that $\poss(\class{\varphi})=\poss(\class{\varphi\land\psi})$, and by the monotonicity of $\poss$, $p\le\poss(\class{\varphi\land\psi})\le\poss( \class{\psi})$.
		
		Axiom \ref{axposs}\ref{axposs7} is valid by the given semantics, and the preservation of validity by the inference rules is proved as in the proof of Theorem \ref{conseqT}.
	\end{proof}
	
	We construct once again the canonical measure by defining for each $\ded_{\Delta_{PsS}}$-maximal set $x\in X_{\Delta_{Ps}S}^D$, and $\varphi::S$, $\poss_x(\vabs{\varphi}_S)=\sup\{q\in[0,1]\st\pposs{\ge q}\varphi\in x\}$. Thus we can prove as in Lemma \ref{propmedcan}:
	
	\begin{lem}\label{propmedcanposs}
		For each $x\in X_{\Delta_{Ps}}^D$
		\begin{enumerate}[\normalfont(1)]
			\item $\poss_x(\vabs{\varphi}_S)<p$ implies $\neg\pposs{\ge p}\varphi\in x$.\label{propmedcanposs1}
			\item $\pposs{\le p}\varphi\in x$ implies $\poss_x(\vabs{\varphi}_S)\le p$.\label{propmedcanposs2}
		\end{enumerate}
	\end{lem}
	
	\begin{thm}\label{canonicalpossmeasurebuenadef}
		$\poss_x$ is a well-defined possibility measure on $\Sigma_S^D$.
	\end{thm}
	\begin{proof}
		As in the proof of Theorem \ref{canonicalmeasurebuenadef}, it follows that $\poss_x(\emptyset)=0$, now using Lemma \ref{modalprinciplesposs}\ref{propps2}, and $\poss_x(X)=1$ follows using the Necessitation Rule and Lemma \ref{modalprinciplesposs}\ref{propps1}.
		
		Let us show that $\poss_x$ satisfies \ref{06072023-3}. Notice that (i) ``if $\pposs{\ge p}\varphi\in x$ then $\pposs{\ge p}(\varphi\lor\psi)\in x$''. Indeed, from the tautology $\varphi\to(\varphi\lor\psi)$, the Necessitation rule gives $\nnec{\ge1}(\varphi\to(\varphi\lor\psi))\in x$. Applying axiom \ref{axposs}\ref{axposs6} it follows that $\pposs{\ge p}\varphi\to\pposs{\ge p}(\varphi\lor\psi)\in x$, so by hypothesis and Detachment we have that $\pposs{\ge p}(\varphi\lor\psi)\in x$. Therefore, from (i) we have that $\sup\{q\st\pposs{\ge q}\varphi\in x\}\le\sup\{q\st\pposs{\ge q}(\varphi\lor\psi)\in x\}$, that is (ii) $\poss_x(\vabs{\varphi}_S)\le\poss_x(\vabs{\varphi\lor\psi}_S)$. Similarly, we can prove that (iii) $\poss_x(\vabs{\psi}_S)\le\poss_x(\vabs{\varphi\lor\psi}_S)$. From (ii) and (iii) we conclude that $\max\{\poss_x(\vabs{\varphi}_S),\poss_x(\vabs{\psi}_S)\}\le\poss_x(\vabs{\varphi\lor\psi}_S)$. 
		
		Assume now, by way of contradiction, that the inequality is strict. Then, there exists $c\in\R$ such that $\max\{\poss_x(\vabs{\varphi}_S),\poss_x(\vabs{\psi}_S)\}<c<\poss_x(\vabs{\varphi\lor\psi}_S)$. In particular $\poss_x(\vabs{\varphi}_S)<c$ and $\poss_x(\vabs{\psi}_S)<c$. By Lemma \ref{propmedcanposs} \ref{propmedcanposs1}, it follows that $\neg\pposs{\ge c}\varphi\in x$, or equivalently $\pposs{<c}\varphi\in x$, and by axiom \ref{axposs}\ref{axposs2} we have that (iv) $\pposs{\le c}\varphi\in x$. Analogously we deduce that (v) $\pposs{\le c}\psi\in x$.
		By axiom \ref{axposs}\ref{axposs5}, it follows  from (iv) and (v) that  $\pposs{\le c}(\varphi\lor\psi)\in x$. On the other hand, from $c<\poss_x(\vabs{\varphi\lor\psi}_S)$ and the contrarreciprocal of Lemma \ref{propmedcanposs} \ref{propmedcanposs2}, it follows that $\pposs{\le c}(\varphi\lor\psi)\notin x$, a contradiction.
		
		Hence, $\max\{\poss_x(\vabs{\varphi}_S),\poss_x(\vabs{\psi}_S)\}=\poss_x(\vabs{\varphi\lor\psi}_S)$, and $\poss_x$ verifies \ref{06072023-3}.
	\end{proof}
	
	\subsection{Completeness for uncertainty polynomial functors}
	Now we can put together the results from the previous sections and give a completeness theorem for the logics of all 
	
	\begin{defi}
		The class of {\em uncertainty polynomial functors} is the one containing the identity functor on $\meas$, the constant functor for each uncertainty space, and is closed under binary products, coproducts, and composition with the functors $\Delta^*, \Delta,\Delta_{Pl}$, and $\Delta_{Ps}$.
	\end{defi}
	
	\begin{exa} The uncertainty polynomial functor $\Delta^*(X\times M)+\Delta_{Ps}(X\times M)$ can be used to represent a system in which different agents choose to represent their beliefs with either upper probability measures or possibility measures.
		
		The delta functors may be nested as needed to model systems analog to those combining probabilities and non-determinism (see \cite{sokolova11probabilistic}). 
	\end{exa}

	\begin{defi}
		For an uncertainty polynomial functor $T$, a $T$-deduction system is a set of relations $\{\ded_S\subseteq$\linebreak[3]$\mathcal{P}(\mathsf{Form}_S)\times\mathsf{Form}_S|S\in\ing T\}$  that is closed under all the inference rules from definitions \ref{defdedsystem},   \ref{defdeductivesystemprob}, \ref{plausAxiomsAndRules}, and \ref{possAxiomsAndRules}. For the assumption rule, we take of course all the axioms (1) through (10) from those definitions.
	\end{defi}
	
	The completeness theorem follows from a construction analogous to the one made in section \ref{seccanonicalspacesandcoalgebras}.
	
	\begin{thm}
		$T$-deduction systems for uncertainty polynomial functors are sound with respect to the coalgebraic semantics. For any Lindenbaum $T$-deduction system $D$, a canonical coalgebra $(X_{Id},\alpha^D)$ can be built. This coalgebra is also final in the category of $T$-coalgebras and proves the completeness of the system $D$.
	\end{thm}
	
	\begin{proof} We already proved soundness for upper probability polynomial functors in Theorem \ref{soundness}, and the same proof works for probability measures. For plausibility and possibility measures, it was proved in Lemmas \ref{plaussoundness} and \ref{possoundness}, respectively.
		
		In order to build the canonical coalgebra for $T$, we need the measurable functions $\rho_{US}:X_{US}\to U(X_S)$ for all the ingredients $S$ of $T$, and $U$ equal to $\Delta^*, \Delta, \Delta_{Pl},$ and $\Delta_{Ps}$. This is done as in the proof of Lemma \ref{measurablefunctions} (4), now using the canonical measures obtained in Theorems \ref{canonicalprobmeasurebuenadef}, \ref{canonicalplausmeasurebuenadef}, and \ref{canonicalpossmeasurebuenadef}, respectively. Thus $\ere_{US}:X_{US}\to US(X_{Id})$ can be defined as $U\ere_S\circ\rho_{US}$, and is measurable as well, leading to the definition, for any Lindenbaum $T$-deduction system $D$, of the canonical coalgebra $(X_{Id},\alpha^D)$, where $\alpha^D:X_{Id}\to T(X_{Id})$ is defined as $\ere_T\circ\rho_{Id}$. This is also the final $T$-coalgebra, as can be seen using the description functions as in Theorem \ref{propertiesdescription} \ref{desfinalcoalgebra}.
		
		Finally for a Lindenbaum $T$-deduction system $D$, the completeness is obtained from the equivalence of $\Gamma\sat_S\varphi$, $\Gamma\sat_S^{\alpha^D}\varphi$, and	 $\Gamma\ded_S^D\varphi$, just as in Corollary \ref{coroComplete}.
	\end{proof}
	
	\bibliographystyle{alphaurl}

\end{document}